**Influence of time, temperature, confining pressure and fluid content on the experimental compaction of spherical grains**

M. Rossi[*1,a], O. Vidal[1], B. Wunder[2], F. Renard[3]

[1] Laboratoire de Géodynamique des Chaînes Alpines, CNRS, OSUG, Université Joseph Fourier, BP 53, 38 041 Grenoble Cedex 9, France
Fax: +33 4 76 51 40 58, E-mail addresses: mrossi@ujf-grenoble.fr, ovidal@ujf-grenoble.fr

[2] GeoForschungsZentrum Potsdam, Division 4.1, Telegrafenberg, 14473 Potsdam, Germany
Fax: +49 331 288 1402, E-mail adress: wunder@gfz-potsdam.de

[3] Laboratoire de Géophysique Interne et de Tectonophysique, CNRS, OSUG, Université Joseph Fourier, BP 53, 38 041 Grenoble Cedex 9, France
Fax: +33 4 76 82 81 01, E-mail address: frenard@obs.ujf-grenoble.fr

* Corresponding author
Phone: +33 (0)4 79 75 94 28
Fax: +33 (0)4 79 75 87 77
E-mail address: Magali.Rossi@univ-savoie.fr
[a] Present address: Environnement Dynamique et Territoires de Montagnes, UMR CNRS 5204, Bât. Belledonne, Université de Savoie, F-73373 Le Bourget du Lac, France.

**Abstract**

26   Theoretical models of compaction processes, such as for example intergranular
27 pressure-solution (IPS), focus on deformation occurring at the contacts between spherical
28 grains that constitute an aggregate. In order to investigate the applicability of such models,
29 and to quantify the deformation of particles within an aggregate, isostatic experiments were
30 performed in cold-sealed vessels on glass sphere aggregates at 200 MPa confining pressure
31 and 350°C with varying amounts of fluid. Several runs were performed in order to investigate
32 the effects of time, fluid content, pressure and temperature, by varying one of these
33 parameters and holding the others fixed. In order to compare the aggregates with natural
34 materials, similar experiments were also performed using quartz sand instead of glass spheres.
35 Experiments with quartz show evidence of IPS, but the strain could not be quantified.
36 Experiments with glass spheres show evidence of several types of deformation process: both
37 brittle (fracturing) and ductile (plastic flow and fluid-enhanced deformation, such as IPS). In
38 experiments with a large amount of water ($\geq$ 5 vol.%), dissolution and recrystallization of the
39 glass spheres also occurred, coupled with crystallization of new material filling the initial
40 porosity. Experiments performed with a fluid content of less than 1 vol.% indicate creep
41 behavior that is typical of glass deformation, following an exponential law. These
42 experiments can also be made to fit a power law for creep, with a stress exponent of
43 $n = 10.5 \pm 2.2$ in both dry and wet experiments. However, the pre-factor of the power law
44 creep increases 5 times with the addition of water, showing the strong effect of water on the
45 deformation rate. These simple and low-cost experiments provide new insights on the
46 rheology of plastically deformed aggregates, which are found in many geological
47 environments, such as partially molten-rocks, pyroclastic deposits or fault gouges during the
48 inter-seismic period.

49

50 **Keywords:** isostatic experiments, glass spheres, ductile deformation, compaction

51

52  **1   Introduction**

53   Ductile deformation usually occurs in the lower crust under high pressure (P) and
54 temperature (T) conditions by crystal plasticity, but can also be efficient under mid-crustal P-
55 T conditions when deformation is accommodated by diffusive mass transfer through a fluid
56 phase (intergranular pressure-solution – IPS – ; e.g., Weyl 1959; Rutter, 1976; Tada & Siever
57 1989; Gratier *et al.* 1999). IPS results from a stress gradient at grain-scale that leads to dissolution and flattening of grain contacts, transport through the contacts to the pores, and precipitation of new phases within the porosity (e.g., Weyl 1959; Tada & Siever 1989). At higher confining pressure, plastic deformation can occur at grain contacts leading to contact flattening. These two ductile deformation mechanisms thus compact and strengthen the initial aggregate, leading to a reduction in porosity and permeability.

Numerous experimental studies have been performed in order to understand the mechanical processes and to quantify the strain of mineral aggregates. High-temperature compaction experiments have been performed in order to investigate plastic deformation in the lower crust and in high-strain zones using both dry and hydrated aggregates (e.g., Rybacki & Dresen, 2000; Xiao *et al.*, 2002; Rutter & Brodie, 2004). Under the P-T conditions of the middle and upper crust, creep experiments were performed in the presence of fluid, including 1) compaction experiments in drained conditions, (e.g., Renton *et al.* 1969; Rutter 1983; Kronenberg & Tullis 1984; Gratier & Guiguet 1986; Rutter & Wanten 2000), 2) compaction experiments with a controlled fluid pressure (i.e. the effective pressure: Niemeijer *et al.* 2002; Niemeijer & Spiers 2002; He *et al.*, 2003), and 3) shearing experiments (Bos & Spiers, 2000; Stünitz & Tullis 2001).

Supplementing the experimental approaches, several numerical and theoretical models have been proposed to explain creep compaction (e.g. Weyl, 1959; Rutter, 1976; Raj & Chyung 1981; Tada & Siever 1986; Tada *et al.* 1987). These models are generally based on the deformation of spherical elements making up an aggregate (e.g., Lemée & Guéguen, 1996; Renard *et al.* 1999, 2000; Gundersen *et al.* 2002; Yasuhara *et al.* 2003). The comparison between models, experiments and nature is not straightforward because compaction experiments are generally conducted with grains of irregular shape, and because models focus on the grain-scale deformation, whereas compaction experiments focus on bulk deformation of the aggregate.

Considering an aggregate in a closed system, subjected to an external hydrostatic pressure, the effective normal stress at the grain contacts depends on the external confining pressure, the pore fluid pressure, and the contacts surface area that increases with compaction. Therefore, the final morphology of the spheres and the size of the contact areas can be used to estimate the local effective stress at the grain contacts at the end of experimental runs. Experiments of different run times involving calibrated spheres instead of grains of irregular shape, conducted in closed systems with different starting solid/fluid ratios could be used to quantify strain at the particle scale and to constrain general strain-stress relations, even without a proper knowledge of the applied stress or strain rate. The aim of the present study was to test this "grain-scale" approach by conducting compaction experiments on an aggregate of calibrated spherical particles, in order to quantify particle strain at the end of the runs by observing the grain contacts using a Scanning Electron Microscope (SEM).

Such experiments could be conducted using very simple experimental equipment, such as isostatic cold-sealed autoclaves or internally-heated apparatus, thus reducing the duration of experiment preparation, and allowing experiments to be performed over a wide range of confining pressure and temperature values. The main disadvantage of such an experimental set-up is the change in effective stress during the experiment, due to an increase in the contact surface area between the spheres with increasing compaction. Therefore, the strain rate and the stress at the contacts are not controlled and vary during the experiment, making it difficult to derive general compaction laws. However, the combination of experiments conducted with the same initial conditions (sphere diameter, water/sphere ratio, temperature, and confining pressure) but different run durations should allow the time-dependent change in strain rate and local stress to be estimated at the contact of the spheres with deformation. To the authors' knowledge, such an approach has not been attempted so far as it requires working with aggregates with constant particle geometry and size. Owing to the difficulties in producing perfectly spherical mineral grains with a small and constant size of about 100 μm, which are required for such experiments, commercial glass spheres had to be used. Moreover, dissolution and precipitation kinetics of glass are higher than in crystallized materials, thus allowing shorter experimental durations.

The present work reports on experiments conducted in closed system conditions varying the time, the solid/fluid ratio, the confining pressure and the temperature. The results open up the possibility of studying rock deformation under isostatic confining pressure, and using glass spheres as an analogous material for silicate minerals. For comparison with natural minerals, similar experiments were conducted with non-spherical quartz grains instead of glass spheres. The results obtained from this very simple experimental approach are used to discuss the creep law of an aggregate made of ductile glass beads and the controlling effect of water content, temperature and pressure on deformation.

## 2 Experimental method

The starting material was an aggregate of 45-90 μm diameter spheres of soda-lime glass (72% $SiO_2$, 14% $Na_2O$, 10% CaO and 3% MgO) from Sandmaster-France. At atmospheric pressure this glass has a softening point at 730°C. According to figure 1, and assuming that viscosity increases with pressure, the glass should behave elastically under

126  stress at T < 450°C. Therefore, and since it is the IPS processes that were of prime interest,
127  most experiments were conducted at 350°C and 200 MPa confining pressure. For comparison
128  with crystallized material, additional experiments were performed with quartz sand (Nemours
129  sand; $d_{50}$=199 μm). In order to investigate the variation in strain with time, several runs were
130  made at the same P-T conditions for durations ranging from 2 hours to 4 weeks. To
131  investigate the effect of temperature at constant pressure, a set of experiments was conducted
132  at 25, 150, 250, 350 and 450°C and a constant pressure of 200 MPa. The effect of pressure
133  was studied at a constant temperature of 350°C, with experiments conducted at 100, 200, 300
134  and 400 MPa (table 1).

135  In all experiments, the glass spheres (or quartz grains) were mixed with about 3 wt.%
136  biotite to investigate the possible enhancement of IPS at the contacts with phyllosilicates (Bos
137  & Spiers, 2000; Rutter & Wanten, 2000; Niemeijer & Spiers, 2002). 100 mg of this was
138  loaded with varying but controlled amounts of fluid into a 3.0 mm outside diameter, 0.2 mm
139  wall thickness and 12 mm long cylindrical gold capsule (figure 2A). The confining pressure
140  minus the pressure of the water in the capsule equals the effective confining pressure To
141  investigate the effect of fluid availability on deformation, various amounts of fluid were thus
142  introduced in the sample: 0, 1 and 5 vol.% double de-ionized water (volume of fluid related to
143  100% volume of glass spheres). These amounts of water give rise to different fluid ($P_f$) and
144  effective ($P_e$) pressures. Assuming a body-centered cubic packing, the starting porosity of the
145  sphere aggregate is about 32%. Considering an initial porosity of 32%, the density of the air +
146  water fluid in the 5 vol.% water experiments was 0.12 g.cm$^{-3}$. From the P-V-T data of
147  Burnham *et al.* (1969) and Keenan *et al.* (1969), this corresponds to a $P_f$ ~ 25 MPa at 450°C.
148  The 0.12 g.cm$^{-3}$ iso-density curve intersects the vapor-liquid equilibrium curve at about 350°C
149  and 17 MPa. Below this temperature, $P_f$ lies on the liquid-vapor curve, i.e. about 10 MPa at
150  300°C and 4 MPa at 150°C. Experiments performed with 20 vol.% of fluid at 400°C would
151  thus have a fluid pressure of about 40 MPa. Therefore, the fluid pressure in the 1 to 5 water
152  vol.% experiments was about 12% of the confining pressure ($P_c$), and Pf < 0.1 x Pc in
153  experiments conducted at T < 400°C. The fluid pressure was higher in experiments conducted
154  with higher amounts of fluid, but it was always less than 0.25 x Pc (maximum value
155  calculated for the 400°C experiment performed with 20 vol.% fluid). These calculations show
156  that the effective pressure was almost the same or close to the confining pressure in most
157  experiments.

158  Gold capsules were used because gold is not reactive to the materials and fluids used
159  for this study. Moreover, gold deforms easily and the external confining pressure is completely transmitted to the enclosed experimental mixture. Once filled, the capsules were welded shut and placed in a water-pressurized horizontal cold-sealed pressure vessel (Tuttle, 1969; figure 2B). The capsules were weighed after each stage of sample preparation to check for any leakage. Confining pressure was measured to within ± 5 MPa, and temperature was measured to within ± 1°C with a chromel-alumel thermocouple located inside the vessel at the contact with the gold capsule. Because of the strong inertia of the heating system and the small size of the capsule, the temperature gradient at the hot end of the autoclave and along the capsule was negligible (Vidal, 1997; Vidal & Durin, 1999).

In each experiment, pressure was applied first, and the sample was then heated up to the desired temperature in about two hours, keeping the pressure constant. At the end of the runs, temperature was decreased to 25°C in less than 15 minutes. The remaining pressure was released when the samples had reached room temperature.

After the runs, the capsules were weighed to check for any water loss and leakage. The capsules were then opened and some material was collected for SEM observation. Except for experiments performed with ≥ 5 vol.% fluid, aggregate cohesion was lost during this procedure, which facilitated observations of the grain contact size and morphology. The run material was deposited on gold-coated copper holders and observed with a Zeiss DSM 962 SEM (GFZ Potsdam, Germany) and a Leica Stereoscan 440 SEM (Université de Savoie, Chambéry, France). Local strain at the grain contacts was calculated from the change in grain geometry measured on the SEM pictures (see part 3.2). The amount of fluid that possibly entered the glass spheres during the experiments was estimated qualitatively from Raman spectroscopy, using a Jobin-Yvon LabRam HR800 (ENS-Lyon, France), conducted directly on the spheres or on double polished thin sections made from epoxy-impregnated samples.

## 3 Results

*3.1 Effect of water content on the morphology of contacts*

The morphology of the grain-grain contacts characterized under the SEM is strongly dependent on the amount of fluid within the sample.

*In dry experiments*, the contacts between the grains showed flat circular morphologies (figure 3A) and the spheres were not bound together. Despite flattening at the contact, there was no significant reduction in porosity after the runs. *In experiments with 1 vol.% fluid*, most spheres were bound together and the contacts showed a flat rim and conchoidal cracking at the core (figure 3B). Numerous contacts showed concentric circles of alternating outgrowths and depressions (figure 3C), and rare stylolite-like contacts were observed, with the spheres both dissolving and penetrating one into another (figure 3D). In both dry and 1 vol.% fluid experiments, the initial porosity was well maintained despite deformation: porosity reduction was too small to be observed under the SEM (figure 3A-B). On the contrary, the initial porosity *in experiments conducted with ≥ 5 vol.% fluid* was completely filled with a microporous matrix having the same composition as the glass spheres. The spheres were all bound together, and their free faces showed recrystallized rims (figure 3E-F). Porosity filling and sphere bonding combined to strengthen the material. The sphere diameter was slightly smaller than that of the starting material, and sphere interpenetration was observed, both these phenomena suggesting strong dissolution of the glass during the experiments. A further increase in the amount of fluid produced the same contact morphology, but also increased the reactivity of glass. Complete recrystallization of the spheres was observed in experiment ENS2 conducted with 15 vol.% fluid (figure 3G).

Raman spectroscopy was used to check the possible incorporation of water in the glass spheres and newly formed material in the porosity. Raman spectra of the initial glass spheres indicated that they were water-free (figure 4Aa). *In experiments with 1 vol.% fluid*, very smooth peaks were detected in the spectral range of water (3500 - 3750 $cm^{-1}$), indicating that if any, only a very small amount of water was absorbed at the rim of the spheres (figure 4Aa). The Raman spectra of the material filling the porosity *in experiments with 5 vol.% fluid* showed well-defined bands at 3550-3750 $cm^{-1}$, 615 and 1100 $cm^{-1}$ (figure 4A), which are characteristic of a hydrous crystalline phase. From comparison of the Raman spectra, this material resembles a Ca-bearing zeolite (figure 5), even though it is Al-free. A profile from this recrystallized material towards and through a glass sphere shows that glass hydration and recrystallization also occur at the rim of the glass spheres (figure 4B). At the contact between two glass spheres, neither glass crystallization nor glass hydration was observed or detected (figure 4B). These observations suggest that glass hydrolysis was associated with recrystallization. Furthermore, the initial macroporosity disappeared and was replaced by a fluid-rich microporous recrystallized matrix. At the sphere contacts, the availability of water was much lower than on the free surfaces. Therefore, no water was incorporated in the glass and no recrystallization occurred.

Regardless of the amount of fluid, evidence of cracking was found in some spheres in all experiments with the cracks spreading from one sphere contact to the next (figure 3H). Cracking increased significantly at low temperature and high confining pressure, but no relationship could be found between the amount of fluid and the amount of cracking. The timing of crack formation remains uncertain. Most cracks probably formed during the very first stage of the experiments during pressure loading, and at the end of the experiments during quenching. However, cracks might also have been formed during the runs.

The occurrence of all the features previously described indicates at least three main responses to stress: ductile deformation in experiments performed with no or a small amount of fluid ($\leq 1$ vol.%), brittle deformation in all runs, and chemical reaction (dissolution-crystallization) in experiments with a large amount of fluid ($\geq 5$ vol.%).

*3.2 Time-dependent change in strain and T, P, water content dependency*

1-D strain was estimated from the change in diameter of the glass beads under compaction (figure 6A). Since the changes in the average sphere diameter due to compaction were negligible, the 1-D strain associated with each contact ($\varepsilon$, shortening of the sphere radius at the contact) can therefore be calculated using the following relation:

$$|\varepsilon| = \frac{d_s - \sqrt{d_s^2 - d_c^2}}{d_s} \quad (1)$$

where $d_c$ is the contact diameter and $d_s$ the associated sphere diameter directly measured for each sphere on the SEM pictures (figure 6A). The size of the glass spheres $d_s$ and the contact size $d_c$ were determined from SEM pictures of the run products. Only unfractured and unbroken contacts were measured in order to estimate ductile strain. For each experiment, about 100 contacts were measured to calculate average contact diameters (Nc in table 1), to ensure that the local strain measured at different grain contacts is representative of the overall strain of the aggregate. Since the confining pressure was isotropic, the strain was also assumed to be isotropic. Equation 1 was only used to estimate the strain in dry experiments and experiments with 1 vol.% fluid for T $\leq$ 350 °C, i.e., experiments for which no glass recrystallization occurred. In experiments with 5 vol.% fluid or with 1 vol.% fluid at 450°C, the material is not loose and just a few sections of contacts could be observed with SEM. It was thus not possible to obtain a good statistical representation of the strain, which would thus be extremely badly constrained.

From equation 1 the finite strain at each contact was calculated at the end of the runs and the strain averaged over a significant number of contacts to obtain the overall strain achieved in each experiment (table 1). Figure 6B shows that in all experiments, most deformation occurred at the very beginning of the experiments, during the first ten hours. Then, deformation quickly slowed down after a few hours. Deformation-time relations were derived by combining the results obtained from different run-time experiments conducted under the same P-T conditions and fluid/solid ratio (table 1, figure 6B). During the first two hours of each experiment, a strong increase in sphere contact area was observed, which implies a strong decrease in stress and strain rate. For this reason, the $\varepsilon = f(t)_{[P,T]}$ relations were fitted with time-dependent power laws using only the strain data at t > 2 hours. A power law exponent value of 0.068 is best constrained by the 1vol.% water experiments at 350°C (figure 6B). In view of the wide scatter of the values obtained in all the experiments, this power exponent was left unchanged to fit the other experiments. For the 150°C experiments, both the wet and dry experiments were fitted using the same $\varepsilon = \alpha\, t^{0.068}$ function (dashed line in figure 6B). Then, discrete differentiation of the ε-time relation data at given P and T values gives the microscopic strain rate (values listed in table 1).

At 350°C, strain was clearly greater in the 1 vol.% fluid experiments than in the dry experiments, whereas the addition of water in the 150°C experiments did not show any significant difference compared to the dry experiments. The strain versus time plots in figure 6B show that strain increased significantly with temperature. This can also be seen in figure 7, which shows a plot of the strain observed after 6 hours in experiments conducted at a pressure of 200 MPa, and temperatures of 25, 150, 250, 350 and 450°C. At T ≥ 250°C strain is enhanced by the addition of water. No quantitative data are available for the 1 vol.% fluid experiment performed at 450°C. Indeed, as observed at 350°C in experiments with 5 vol.%, the spheres chemically reacted with the fluid and the porosity was filled with new material.

A set of experiments with 6-hour run time was also performed at various confining pressures (100, 200, 300, 400 MPa) at 350°C (figure 8). As expected from the previous ε-T results after 6 hours, experiments containing 1 vol.% fluid showed a higher strain than dry experiments at all pressures. Radial fracturing spreading from the contacts increases with confining pressure (figure 3H), so that the number of contacts between unbroken spheres is lower in experiments with high rather than low confining pressure. However, figure 8 shows that, for a constant water content, the average strain increases linearly with increasing confining pressure.

All these observations indicate that strain increases with temperature, confining pressure, and the addition of even very small amounts of water. The similarity observed at 350°C in 5 vol.% fluid experiments and at 450°C in 1 vol.% fluid experiments, and the fact that, at 150°C, dry and 1 vol.% fluid experiments have similar strain-time variations, suggest that the effect of water was thermally activated. In contrast, the effect of pressure was similar for the dry and 1vol.% fluid experiments. The observed increase in strain was thus simply related to an increase in local stress at the contact between the spheres.

*3.3 Effect of phyllosilicates*

3 wt.% biotite was introduced in the starting mixture, in order to investigate whether or not the presence of mica enhances deformation as discussed in the literature (Rutter 1983; Tada & Siever, 1989; Dewers & Ortoleva 1995; Hickman & Evans 1995; Bos *et al.* 2000; Rutter & Wanten 2000; Renard *et al.* 2001; Niemeijer & Spiers 2002). Glass sphere – mica contacts were scarce and difficult to identify because the samples lost their cohesion when opening the capsule. In dry experiments, the mica deformed mechanically by kinking and indentation (figure 9A). In experiments with 1 vol.% fluid, the mica was also kinked (figure 9B) and the spheres were slightly dissolved at their contacts (figure 9C). The mica could also be dissolved when a fine flake was sandwiched between two spheres. In experiments with 5 vol.% fluid, it was not possible to make any systematic direct observations of the mica because the mica flakes were closely sandwiched inside an aggregate of very compact recrystallized glass spheres. Some spheres were occasionally truncated at their contacts with mica (figure 9D), but it was hardly possible to quantify any strain or dissolution of these spheres. If at all, the dissolution of glass spheres at the contact with mica was low, and mica also dissolved.

Even though glass sphere dissolution was evidenced at the contact with the mica, it was impossible to quantify the amount of dissolution and the resulting strain related to the presence of mica. However SEM observations seemed to indicate that the effect of mica on glass sphere dissolution was mostly limited to spheres directly in contact with the mica flakes.

*3.4 Comparison with natural materials*

In order to compare the results obtained for glass spheres with natural materials, several experiments were carried out using natural quartz under the same P-T and water content conditions as used with the glass beads (table 1). The initial grains already presented some features of previous dissolution, but the surface of the grains was relatively smooth. After a few weeks, micro-stylolites developed at the contact between quartz grains, forming rough surfaces, while free faces only showed dissolution patterns (figure 10). These observations indicate that IPS was efficient under isostatic compaction even with very small amounts of water. However, the unknown and variable initial shape of the quartz grains precluded any quantification of the strain of individual grains, as done for the glass spheres.

## 4 Discussion

The aim of this study was to quantify the strain of the particles of an aggregate using a simple experimental design (isostatic confining pressure, constant temperature), for various P-T-t conditions and amounts of fluid. Deformation processes and the heterogeneity of the local stresses are discussed first. The experimental results are then compared with glass creep according to an exponential law and with the creep of natural minerals according to a power law. Finally, the implications of this experimental study are discussed for several geological fields where fluid-enhanced deformation under high P-T is relevant.

### *4.1 Deformation processes*

Experiments were performed under P-T conditions suitable for IPS deformation in natural silica-rich rocks, but the experimental results obtained indicate that at least three main deformation processes occurred in the simple system used, involving only glass spheres + 3 wt.% mica and pore water:

- *Brittle deformation* (cracking) occurred in all experiments, but the level of deformation depends primarily on the confining pressure and to a lesser extent on temperature.

- *Ductile deformation* was also observed in all experiments, illustrated in the dry experiments by the development of flat contacts due to plastic flow. This deformation process is very rapid and occurred at the very beginning of all experiments. Figure 7 shows that plastic flow was thermally activated, which is consistent with a decrease in effective glass viscosity with increasing temperature (figure 1). Plastic flow also occurred in the experiments conducted with water, as suggested by the flat outer rings bordering the contacts between the spheres in these experiments. Although glass hydration was not evidenced by Raman spectroscopy in the 1vol.% fluid experiments, the increase in strain with the addition of water is likely due to the decrease in effective glass viscosity resulting from partial hydration of the glass. Water affects glass by decreasing the connectivity of the network according to the following chemical reaction:

$$\text{—Si—O—Si—} + H_2O \rightarrow \text{—Si—OH HO—Si—} \qquad (2)$$

This reaction leads to the replacement of covalent chemical bonds by hydrogen bonds, thereby inducing a reduction in the effective glass viscosity. The observed enhancement of recrystallization with 1 vol.% fluid at T > 350°C suggests that the hydration reaction of glass was thermally activated. However, another deformation process occurred in the experiments with water. Given the shape of the inside surface of the contacts and the fact that spheres were often bonded together it would seem that not all the water entered the glass spheres, and that fluid-driven deformation, such as IPS, might also have been active in the 1 vol.% water experiments. Fluid-driven deformation was clearly less efficient than plastic flow at the beginning of the experiments. However, it became the dominant deformation mechanism in experiments lasting more than two weeks, once the contact between the spheres reached a critical size above which the local stress was too low for plastic deformation. In any case, the low amount of fluid was not a limiting factor for IPS, since it proved to be effective in the quartz bearing experiments with 1 vol.% water. The observations made in these experiments also suggest that, in the presence of a small amount of water, there is a critical contact size above which sintering occurred. This effect strengthened the aggregate and acted against further ductile deformation.

- *Chemical solid-fluid reactions* occurred at 450°C even for low fluid content and at 350°C in the 5 vol.% fluid experiments. Under these conditions, the glass spheres crystallized and also dissolved on their free faces. This is reflected by the occurrence of reaction rims on free faces and by the presence of a new phase in the porosity (figures 3E-G). Assuming that the initial porosity of the aggregates was about 32% (based on a body-centered cubic packing), 5 vol.% and 1 vol.% of water filled respectively 14% and 3% of the porosity. The extent of the reaction was surprisingly high in view of the low amount of fluid involved in the experiments and considering that the transport of elements from the spheres toward the porosity has to take place by diffusion in a thin fluid layer confined at the grain contacts. Solid-fluid chemical reactions evidenced at 350°C in 5 vol.% experiments and at 450°C in 1 vol.% experiments thus indicate that dissolution and precipitation kinetics, as well as transport kinetics in the fluid layer, were very high. The difference in experimental conditions between experiments showing almost no precipitation in the porosity (1 vol.% fluid at 350°C) and those showing a porosity almost completely filled with newly crystallized products (5 vol.% fluid at 350°C and 1 vol.% fluid at 450°C) indicates that the extent of fluid-glass interactions was closely dependent on both temperature and amount of fluid.

## 4.2 Heterogeneity of local stress in an aggregate

Local normal stresses applied at the sphere contacts ($\sigma_c$) at the end of the experiment were calculated from the size of the contact and the applied confining pressure using:

$$\frac{\sigma_c}{d_s^2} = \frac{4\sigma}{\pi d_c^2} \tag{3}$$

where $\sigma$ is the confining pressure (MPa), $d_s$ the sphere diameter (m) and $d_c$ the contact diameter (m). Local stresses were calculated for each contact and then averaged over the sample. According to the strain estimates given in table 1, porosity changes related to compaction were less than 2% of the initial porosity, so that changes in fluid pressure during compaction can reasonably be neglected. Furthermore, in experiments with 5 vol.% fluid with an initial 30% porosity, the pore pressure developed up to 350°C would be only up to a maximum of about 10 MPa, whereas at 450°C, under supercritical conditions it would be about 40 MPa. Consequently, the fluid pressure is much lower than the local stress at the grain contacts (table 1) and can therefore be neglected when estimating the effective stress at grain contacts according to equation 3.

Except for experiments performed at 400 MPa, cracking was only evidenced at a few contacts and was therefore observed to be a secondary deformation mechanism. Equation 3 is thus only suitable for calculating the local normal stress at plastically deformed contacts that were not fractured.

In all experiments, the dispersion of contact diameters is 2 to 7% of the average diameter (table 1), showing that the strain was heterogeneously distributed at grain scale in the samples. The calculated local stress dispersion is 5 to 30% of the averaged local stress, with most experiments having a lower heterogeneity with a dispersion of 5 to 15%. Even if these heterogeneities were observed at small scale, they average in the volume and the deformation at the sample scale was homogeneous. Consequently, using a sufficient number of microscopic measurements, a macroscopic constitutive flow law could be calculated for the aggregate.

*4.3   Comparison with creep of glass*

The relative contributions of the three deformation mechanisms mentioned in part 4.1 may have changed during the experiments, but they could have also been active at the same time. However, the results obtained indicate that strain was mostly accommodated by plastic flow, at least during the dry experiments and the first stage of the experiments with added water. Plastic deformation in glass cannot occur by the movement of dislocations as in crystallized material. Instead, deformation occurs by the rearrangement of groups of 10 to 100 atoms. The macroscopic strain rate $\dot{\varepsilon}$ of plastic deformation in an amorphous material can be described by the following equation (Spaepen, 1981; Heggen *et al.*, 2004):

$$\dot{\varepsilon} = 2 c_f k_f \frac{B}{\Omega} \sinh(\frac{B}{2kT}\sigma) \tag{4}$$

426  where $\sigma$ is the external pressure, $k_f$ is a rate constant, $\Omega$ is the atomic volume, $c_f$ is the
427  concentration of defects, $k$ is the Boltzmann constant, $T$ is the temperature, and $B$ is the
428  activation volume. When $\frac{B}{2kT}\sigma$ is high, $\sinh\left(\frac{B}{2kT}\sigma\right)$ can be simplified to $\frac{1}{2}\exp\left(\frac{B}{2kT}\sigma\right)$
429  and the strain equation 4 can be rewritten as a more classical exponential law.

430 $$\dot{\varepsilon} = AB\exp(\frac{B}{2kT}\sigma) \tag{5}$$

431  with $A = \frac{2c_f k_f}{\Omega}$.

432      A plot of $\dot{\varepsilon}$ as a function of the local normal stress at the contact (figure 11) shows a
433  linear trend on a linear-log scale. Therefore, the experimental data can be reasonably
434  explained by an exponential law (equation 5). The slope of the $\ln\dot{\varepsilon}$-$\sigma$ lines in figure 11
435  corresponds to the values of $B/2kT$, which are respectively $2.69 \pm 1.16 \times 10^{-9}$ and $3.64 \pm 1.96$
436  $\times 10^{-9}$ for dry and wet experiments at 350°C. From these values, activation volumes $B = 46 \pm$
437  20 Å$^3$ and $B = 63 \pm 32$ Å$^3$ are obtained for the dry and wet experiments (table 2). In view of
438  the very wide scatter of the $\ln\dot{\varepsilon}$-$\sigma$ experimental data at 150°C, it was not possible to calculate
439  an activation volume from these data, so it was assumed that it was the same as at 350°C.
440      The pre-exponential factors derived from figure 11 for both the dry and wet
441  experiments at 350°C are: $AB_{dry} = 1.91 \pm 1.32 \times 10^{-12}$ s$^{-1}$ and $AB_{1vol\%} = 4.16 \pm 7.06 \times 10^{-13}$ s$^{-1}$.
442  For the 150°C experiments, a value of $AB = 5.69 \pm 41.03 \times 10^{-15}$ s$^{-1}$ was obtained for both the
443  dry and wet experiments, when using $B = 46 \pm 29$ Å$^3$. The increase in $AB$ with temperature
444  corresponds to an increase in $A$ from about $10^{-16}$ Å$^{-3}$.s$^{-1}$ at 150°C to about $10^{-14}$-$10^{-15}$ Å$^{-3}$.s$^{-1}$ at
445  350°C in both the wet and dry experiments respectively. These variations quantify the joint
446  effect of temperature and water on the rate constant $k_f$ in equation 4.
447      Figure 11 gives $B_{1vol\%} > B_{dry}$ and $A_{1vol\%} < A_{dry}$ (table 2). The presence of fluid in the
448  porosity and in the contact zone between two glass spheres is likely to affect the rate constant
449  $k_f$, while $c_f$ and $\Omega$ remain constant. Therefore, figure 11 suggests that $k_{f\,1vol\%} < k_{f\,dry}$. However,
450  these experiments indicate that strain and strain rate are lower in dry experiments than in
451  experiments performed with 1vol.% fluid, suggesting that the rate constant $k_f$ is larger in the
452  wet experiments. Assuming that the activation volume B is constant in both the dry and wet
453  experiments implies that $A_{1vol\%} > A_{dry}$ and $k_{f\,1vol\%} > k_{f\,dry}$. Indeed, if as shown in figure 11, the
454  best fit of the $\ln\dot{\varepsilon}$-$\sigma$ data obtained at 350°C in the wet experiments is plotted using a fixed $B =$
455  46 Å$^3$ (as obtained from the dry experiments), $AB_{1vol\%} = 7.11 \pm 4.91 \times 10^{-12}$, which 456  corresponds to $A_{1vol\%}$ = 1.54 ± 1.06 x 10$^{-13}$ Å$^{-3}$.s$^{-1}$. This value of $A$ is now higher than that
457  obtained for the dry experiments ($A_{dry}$ = 4.06 x 10$^{-14}$ Å$^{-3}$.s$^{-1}$), which is consistent with an
458  increase in the rate constant $k_f$ resulting from glass hydration. It is therefore likely that the
459  activation volumes of dry and hydrated glass are similar.

460  An interesting result is that the effective glass viscosity estimated as $\eta_{eff} = \sigma/\dot{\varepsilon}$ lies in
461  the range 10$^{15}$ - 10$^{18}$ Pa.s. These values are in agreement with the extrapolation of the
462  viscosity of soda-lime glass measured at high temperature to the temperature of the
463  experiments described here (figure 1). Furthermore, the effective glass viscosity is higher in
464  dry than in wet experiments and in the 150°C experiments than those at 350°C (table 1).

465

466  *4.4 Comparison with power-law creep*
467  Deformation of crystalline material at moderate temperature involves dislocation
468  creep, which is well described by a power-law. As mentioned above, glass deformation does
469  not involve the movement of dislocation and may be described by exponential creep laws.
470  However, in view of the large variation in strain rate with varying contact stresses (more than
471  a factor 100 increase in $\dot{\varepsilon}$ is observed when σ is doubled, see figure 11) and experimental
472  data scatter, it is reasonable to describe these experiments using a power law relationship.
473  Such a constitutive law is widely used for rocks subjected to stress and strain rates
474  corresponding to those of the crust (Poirier, 1985):

475  $$\dot{\varepsilon} = A_1 \exp\left(-\frac{E}{RT}\right)\sigma^n \qquad (6)$$

476  with A$_1$ and n being the power law parameters (A$_1$ in Pa$^{-n}$s$^{-1}$), E the activation energy
477  (kJ.mol$^{-1}$), R the gas constant (kJ.mol$^{-1}$K$^{-1}$) and T the temperature (K). The Arrhenius term
478  accounts for the increase in strain rate with temperature (figure 7). The validity of the power
479  law for the conditions of the experiments described here is demonstrated in figure 12, which
480  shows that the *log($\dot{\varepsilon}$)-log($\sigma$)* data for the different sets of experiments can be reasonably
481  fitted by a straight line.

482  The stress exponent and pre-exponential factors in equation 6 can be obtained from the
483  slope and the intercept of these plots. Figure 12 shows that the dry and wet experiments can
484  be fitted with the same stress exponent, *n* = 10.5 ± 2.2 at 350°C. These values are much
485  higher than those generally obtained for crystalline material deformed by dislocation creep at
486  similar conditions, *n* = 2 to 4 for quartzite (Kronenberg & Tullis, 1984 and references therein;
487  Poirier, 1985; Koch *et al.*, 1989) and *n* = 5-8 for salt (Poirier, 1985). However, the high stress

exponent deduced from the experiments is in reasonable agreement with those reported by Kingery *et al.* (1976) for glass (from 4 to 20) or for plasticine at room temperature (up to $n = 7.3$; Zulauf & Zulauf, 2004). Kronenberg & Tullis (1984) evidenced a decrease of the stress exponent with fluid availability, whereas this study show that both dry and wet experiments have similar stress exponent with similar uncertainties. However, one should note that the pre-factor $A_1 exp(-E/RT)$ of the power-law creep increases 5 times with the addition of water, from $1.50 \pm 1.04 \times 10^{-108}$ Pa$^{-10.5}$s$^{-1}$ in dry experiments to $8.01 \pm 5.54 \times 10^{-108}$ Pa$^{-10.5}$s$^{-1}$ in wet experiments. This evolution shows the strong enhancing effect of water on deformation rate.

The activation energy can be calculated using the intercepts ($A_1 \exp(-E/RT)$ in equation 6) of the $log(\dot{\varepsilon}) = f(log(\sigma))$ fits obtained for the 150°C and 350°C experiments with the same stress exponent. Since the same $\varepsilon = f(t)_{[P,T]}$ function was used to fit the dry and wet experiments at 150°C (dashed line in figure 12), different activation energies are obtained when using the dry or wet experiments at 350°C. Values of $E = 78$ kJ.mol$^{-1}$ and 96 kJ.mol$^{-1}$ were obtained when using the dry and wet experiments respectively, assuming $A_1$ is independent of temperature. $A_1$ calculated values are thus $4.82 \pm 42.16 \times 10^{-102}$ Pa$^{-10.5}$s$^{-1}$ for dry glass and $8.88 \pm 77.73 \times 10^{-100}$ Pa$^{-10.5}$s$^{-1}$ for hydrated glass. For plastic processes such as dislocation creep, the availability of water accounts for a decrease in activation energy from 300 kJ.mol$^{-1}$ in vacuum-dried quartzite to 220-170 kJ.mol$^{-1}$ in "as is" samples down to 120-150 kJ.mol$^{-1}$ in wet quartzites, and for a variable increase in the pre-exponential parameter $A_1$ (Kronenberg & Tullis, 1984 and references therein). The increase in the pre-exponential parameter $A_1$ observed in this study is thus consistent with observations of dislocation creep of quartzite (Kronenberg & Tullis, 1984 and references therein; Koch *et al.*, 1989). Activation energies estimated here for plastic deformation of amorphous materials are 1.5-2 times lower than those estimated for crystalline materials, but they are similar to those of intergranular pressure-solution (73 kJ.mol$^{-1}$ Dewers & Hajash, 1995). The increase of the activation energy from wet to dry experiments is opposite to the variations reported in the literature for crystalline material. However, in view of the large uncertainties on the estimated activation energies, this apparent variation is not significant.

*4.5 Comparison with natural materials*

Experiments with glass spheres indicate a complex deformation process involving mostly plastic flow. Plastic deformation of silica minerals occurs at high pressure and temperature, and can be enhanced in the presence of fluid due to chemical reactions. The simple low-T experiments on glass spheres described here can thus be considered as a first order analogue for the macroscopic study of silicate deformation at higher P-T conditions.

Rock softening due to the presence of fluid is mostly evidenced in the continental crust in ductile shear zones where fluid circulation coupled with high-grade deformation changes the rock rheology (Boullier & Guéguen, 1975; Behrmann & Mainprice, 1987; Martelat *et al.*, 1999; Zulauf *et al.*, 2002; Rosenberg & Stünitz, 2003). These experiments thus have implications in several geological fields where deformation is coupled with fluid-rock interaction under high P-T conditions.

This study shows that despite the competition of several deformation mechanisms, stresses are mostly accommodated by plastic deformation mechanisms in the first-stage of compaction. However, due to the increase in contact surface area with time, the local stresses and resulting plastic deformation efficiency decrease. In the long term, the local stresses are no longer high enough for plastic deformation to be efficient and, as a result, plastic-dominant deformation is replaced by IPS-dominant deformation. Similar coupling and long-term change from cataclastic flow in a short-time scale (seismic event) to plastic and IPS deformations at a longer time-scale (inter-seismic event) has been evidenced in fault gouges in association with faulting and fluid circulation (Gratier *et al.*, 2003; Boullier *et al.*, 2004). Similar experiments to those reported in this study could thus give insights into the interactions between fluid circulation and the mechanical behavior and evolution of fault gouge.

High-temperature deformation experiments indicate that creep is enhanced in the presence of partial melting (Dell'Angelo *et al.*, 1987; Scheuvens, 2002; Garlick & Gromet, 2004). A mechanism change is observed under the same P-T conditions from dislocation creep to a process involving mass transfer by diffusion in the presence of a small amount of melt (< 20% melt, Dell'Angelo *et al.*, 1987; Rosenberg & Handy, 2000; Rosenberg, 2001). The strain rate is higher in the presence of melt because films of melt localized within the grain contact enhance the diffusion of dissolved material to the pores, a process equivalent to IPS. In these experiments, the presence of water in the porosity has the same effect as the presence of small amounts of melt in the porosity of HT rocks (< 20% melt). The use of glass spheres may thus allow analogous melt experiments to be performed using a simple experimental design as well as experiments at lower P-T conditions than usual melt experiments. The main advantage of this experimental set-up is that the initial aggregate geometry can be deconvoluted from the deformation state.

Another field of application for similar experiments such as those conducted here is the compaction of pyroclastic deposits. Deformation experiments of glass sphere aggregates at relevant P-T and water content conditions should provide the necessary data for modeling the deformation of pyroclastic material (Quane *et al.*, 2004; Quane & Russel, 2005; Roche *et al.*, 2005). Quane *et al.* (2004) conducted compaction experiments with glass beads under constant uniaxial strain rate as an analogue for the welding of pyroclastic material. As a complement to such experiments, the approach described here provides another framework for studying more realistic natural situations. The data from these experiments show that hydrated pyroclastic deposits would compact more than dry deposits because of the increase in plasticity due to glass hydration. Glass hydration and associated rheological changes thus increase the compaction rate and allow compaction at lower temperature than for dry materials. Furthermore, these experiments provide new constraints on the rheology of soda-lime glass, which is used for analogue modeling of volcanic processes (Quane *et al.*, 2004; Quane & Russell, 2005).

## 5   Conclusion

Isostatic compaction experiments were performed on quartz and glass sphere aggregates under mid-crustal P-T conditions (200 MPa and 350°C) in order to quantify grain-scale strain and to make a comparison with theoretical models of ductile deformation in granular analogues of rocks. The main advantage of this approach is that, using low-cost pressure vessels, more than sixty deformation experiments could be performed. This made it possible to test a wide range of confining pressures, temperatures, and amounts of fluid.

Under the experimental P-T conditions, quartz and glass spheres do not deform by the same process. Natural materials show evidence of intergranular pressure-solution (IPS), while experiments with glass spheres show evidence of both, brittle and ductile deformation processes. Even though experiments using glass spheres show little evidence of IPS, ductile strain by grain contact plastic flattening can be significant in systems subjected to isostatic pressure conditions. However, IPS is an efficient deformation mechanism in the quartz aggregates experiments in the presence of 1 to 3 vol.% water and in the absence of strong external deviatoric stress.

Dry experiments show that glass spheres accommodate stress concentrations by plastic strain. The addition of water in the porosity allows the local stresses to be accommodated by additional deformation mechanisms. The experiments described show that fluid-glass interactions can be very high, even with very low amounts of fluid at T > 250°C.

Below this temperature, the influence of water on strain decreases rapidly, and no influence was found for long-duration experiments lasting up to two weeks at 150°C with a low fluid content. In contrast, a very high reactivity of glass with water was observed at 450°C, leading to the precipitation of new material filling all the porosity. The same observations were also made at 350°C, but in the presence of a larger (but still very low) amount of water. With a large amount of fluid, chemical processes are dominant due to very rapid kinetics of dissolution-transport-precipitation of the glass.

In dry and wet experiments, deformation of a glass sphere aggregate is accommodated mostly by plastic flow that can be described by either an exponential or a power law for the range of experimental conditions. With power-law creep, the presence of a small amount of water does not affect the stress exponent (n = 10.5 in both the dry and 1 vol.% experiments at 350°C), but leads to a decrease in the pre-factor.

This study clearly highlights the key effect of water, and to a lesser extent temperature, in rock softening. The simple experiments presented in this study can thus be used as a first approximation of natural silicate rocks deformed in the presence of fluid at high pressure and temperature, like for instance the deformation of partially molten-rocks, the compaction of pyroclastic deposits or the change in deformation mechanisms in a fault zone during the inter-seismic phase.

**Acknowledgements**

This work was supported by Dyethi-INSU, STREP-PCRD6 and PROCOPE programs. The authors are grateful to R. Schulz and H. Steigert for technical assistance in the hydrothermal laboratory, and to J. Herwig and H. Kemnitz for their technical assistance during SEM observations. Thanks are also due to J.P. Gratier and E. Lewin for interesting discussions, and to A. Niemeijer, E. Rutter, and an anonymous reviewer for their critical and constructive reviews.

**References**

Behrmann, J.H. and Mainprice, D., 1987. Deformation mechanisms in a high-temperature quartz-feldspar mylonite: evidence for superplastic flow in the lower crust. *Tectonophysics* **140**(2), 297-305.

Bos, B., Peach, C.J. and Spiers, C.J., 2000. Frictional-viscous flow of simulated fault gouge caused by the combined effects of phyllosilicates and pressure solution. *Tectonophysics* **327**(3-4), 173-194.

Bos, B. and Spiers, C.J., 2000. Effect of phyllosilicates on fluid-assisted healing of gouge-bearing faults. *Earth Planet. Sci. Lett.* **184**(1), 199-210.

Boullier, A.M. and Guéguen, Y., 1975. SP-mylonites: origin of some mylonites by superplastic flow. *Contrib. Mineral. Petrol.* **50**, 93-104.

Boullier, A.M., Fujimoto, K., Ito, H., Ohtani, T., Keulen, N., Fabbri, O., Amitrano, D., Dubois, M. and Pezard, P., 2004. Structural evolution of the Nojima fault (Awaji Island, Japan) revisited from the GSJ drill hole at Hirabayashi. *Earth Planets Space* **56**, 1233-1240.

Burnham, C.W., Holloway, J.R. and Davis, F., 1969. The specific volume range of water in the range 1000 to 8900 bars, 20° to 900°C. *Am. J. Sci.* **267-A**, 70-95.

Dell'Angelo, L.N., Tullis, J. and Yund, R.A., 1987. Transition from dislocation creep to melt-enhanced diffusion creep in fine-grained granitic aggregates. *Tectonophysics* **139**, 325-332.

Dewers, T. and Ortoleva, P., 1995. Influences of clay minerals on sandstone cementation and pressure solution. *Geology* **19**, 1045-1048.

Dewers, T. and Hajash, A., 1995. Rate laws for water-assisted compaction and stress-induced water-rock interaction in sandstones. *J. Geophys. Res.* **100**(B7), 13093-13112.

Garlick, S.R. and Gromet, L.P., 2004. Diffusion creep and partial melting in high temperature mylonitic gneisses, hope Valley shear zone, New England Appalachians, USA. *J. metamorphic Geol.* **22**(1), 45-62.

Gratier, J.P. and Guiguet, R., 1986. Experimental pressure solution-deposition on quartz grains: the crucial effect of the nature of the fluid. *J. Struct. Geol.* **8**(8), 845-856.

Gratier, J.P., Renard, F. and Labaume, P., 1999. How pressure solution creep and rate fracturing processes interact in the upper crust to make it behave in both a brittle and viscous manner. *J. Struct. Geol.* **21**(8-9), 1189-1197.

Gratier, J.P., Favreau, P. and Renard, F., 2003. Modeling fluid transfer along California faults when integrating pressure solution crack sealing and compaction processes. *J. Geophys. Res.* **108** (B2), doi:10.1029/2001JB000380.

Gundersen, E., Dysthe, D.K., Renard, F., Bjørlykke, K. and Jamtveit, B., 2002. Numerical modelling of pressure solution in sandstone, rate-limiting processes and the effect of clays. In: de Meer, S., Drury, M.R., de Bresser, J.H.P. and Pennock, G.M., 2002. *Deformation Mechanisms, Rheology and Tectonics: Current status and future perspective*. Geological Society, London, Special Publications, 200, 41-60.

He, W., Hajash, A. and Sparks, D., 2003. Creep compaction of quartz aggregates: effects of pore fluid flow – a combined experimental and theoretical study. *Am. J. Sci.* **303**. 73-93.

Heggen, M., Spaepen, F. and Feuerbacher M., 2004. Plastic deformation of Pd41, $Ni_{10}Cu_{29}P_{20}$

bulk metallic glass. *Mat. Sci. Engin.* **375-377**, 1186-1190.

Hickman, S.H. and Evans, B., 1995. Kinetics of pressure solution at halite-silica interfaces and intergranular clay films. *J. Geophys. Res.* **100** (B7), 13113-13132.

Keenan, J.H., Keyes, F.G., Hill, P.G. and Moore, J.G., 1969. *Steam Tables. Thermodynamic Properties of Water Including Vapor, liquid and Solid Phases.* John Wiley and Sons, Inc. 156 p.

Kingery, W.D., Bowen, H.K. and Uhlman, D.R., 1976. *Introduction to ceramics, 2$^{nd}$ Edition.* John Willey & Sons, New York, 1032p.

Koch, P.S., Christie, J.M., Ord, A. and George R.D., Jr, 1989. Effect of water on rheology of experimentally deformed quartzite. *J. Geophys. Res.* **994**(B10), 13975-13996.

Kronenberg, A.K. and Tullis, J., 1984. Flow strengths of quartz aggregates: grain size and pressure effects due to hydrolytic weakening. *J. Geophys. Res.* **89**(B6), 4281-4297.

Lemée, C. and Guéguen, Y., 1996. Modelling of porosity loss during compaction and cementation of sandstones. *Geology* **24**, 875-878.

Martelat, J.E., Schulmann, K., Lardeaux, J.M., Nicollet, C. and Cardon, H., 1999. Granulite microfabrics and deformation mechanisms in southern Madagascar. *J. Struct. Geol.* **21**(6), 671-687.

Niemeijer, A.R. and Spiers, C.J., 2002. Compaction creep of quartz-muscovite mixtures at 500°C: preliminary results on the influence of muscovite on pressure-solution. In: de Meer, S., Drury, M.R., de Bresser, J.H.P. and Pennock, G.M., 2002. *Deformation Mechanisms, Rheology and Tectonics: Current status and future perspective*. Geological Society, London, Special Publications, 200, 61-71.

Niemeijer, A.R., Spiers, C.J. and Bos B., 2002. Compaction creep of quartz sand at 400-600°C: experimental evidence for dissolution-controlled pressure solution. *Earth Planet. Sci. Lett.* **195**, 261-275.

Poirier, J.P., 1985. *Creep of crystals: high temperature processes in metals, ceramics and minerals*. Cambridge University Press, Cambridge, 260p.

Quane, S.L., Russell, J.K. and Kennedy, L.A., 2004. A low-load, high-temperature deformation apparatus for volcanological studies. *Am. Min.* **89**, 873-877.

Quane, S.L. and Russell, J.K., 2005. Welding: insights from high-temperature analogue experiments. *J. Volcanol. Geotherm. Res.* **142**, 67–87.

Raj, R. and Chyung,C.K., 1981. Solution-precipitation creep in glass ceramics. *Acta Metallurgica* **29**, 159-166.

Renard, F., Park, A., Ortoleva, P. and Gratier, J.P., 1999. An integrated model for transitional

pressure solution in sandstones. *Tectonophysics* **312**, 97-115.

Renard, F., Gratier, J.P. and Jamtveit, B. 2000. Kinetics of crack-sealing, intergranular pressure solution, and compaction around active faults. *J. Struct. Geol.* **22**(10), 1395-1407.

Renard, F., Dysthe, D., Feder, J., Bjørlykke, K. and Jamtveit, B., 2001. Enhanced pressure solution creep rates induced by clay particles: experimental evidence in salt aggregates. *Geophys. Res. Lett.* **28** (7), 1295-1298.

Renton, J.J., Heald, M.T. and Cecil, C.B., 1969. Experimental investigation of pressure solution of quartz. *J. Sediment. Petrol.* **39**(3), 1107-1117.

Roche, O., Gilbertson, M.A., Phillips, J.C. and Sparks, R.S.J., 2005. Inviscid behaviour of fines-rich pyroclastic flows inferred from experiments on gas-particle mixtures. *Earth Planet. Sci. Lett.* **240**(2), 401-414.

Rosenberg, C.L., 2001. Deformation of partially molten granite: a review and comparison of experimental and natural case studies. *Int. J. Earth Sciences* **90**(1), 60-76.

Rosenberg, C.L. and Handy, M.K., 2000. Syntectonic melt pathways during simple shearing of a partially molten rock analogue (norcamphor-Benzamide). *J. Geophys. Res.* **105**(B2), 3135-3149.

Rosenberg, C.L. and Stünitz, H., 2003. Deformation and recrystallisation of plagioclase along a temperature gradient: an example from the Bergell tonalite. *J. Struct. Geol.* **25**, 389-408.

Rutter, E.H., 1976. The kinetics of rock deformation by pressure-solution. *Phil. Trans. R. Soc. Lond.* **283**, 203-219.

Rutter, E.H., 1983. Pressure solution in nature, theory and experiment. *J. Geol. Soc. London* **140**, 725-740.

Rutter, E.H. and Wanten, P.H., 2000. Experimental study of the compaction of phyllosilicate-bearing sand at elevated temperature and with controlled pore water pressure. *J. Sedim. Res.* **70**(1), 107-116.

Rutter, E.H. and Brodie, K.H., 2004. Experimental grain-size sensitive flow of hot-pressed Brazilian quartz aggregates. *J. Struct. Geol.* **26**, 2011-2023.

Rybacki, E. and Dresen, G., 2000. Dislocation and diffusion creep of synthetic anorthite aggregates. *J. Geophys. Res.* **105**(B11), 26017-26036.

Scheuvens, D., 2002. Metamorphism and microstructures along a high-temperature metamorphic field gradient: the north-eastern boundary of the Královský hvozd unit (Bohemian Massif, Czech Republic*). J. metamorphic Geol.* **20**, 413-428.

Spaepen, F., 1981. Defects in amorphous materials. Les Houches Lectures XXXV on Physics of Defects, North Holland Press, Amsterdam, 133-174.

Stünitz, H. and Tullis, J., 2001. Weakening and strain localization produced by syn-deformational reaction of plagioclase. *Int. J. Earth Sciences* **90**(1), 136-148.

Tada, R. and Siever, R., 1986. Experimental knife-edge pressure solution of halite. *Geochim. Cosmochim. Acta* **50**(1), 29-36.

Tada, R., Maliva, R. and Siever, R., 1987. A new mechanism for pressure solution in porous quartzose sandstone. *Geochim. Cosmochim. Acta* **51**, 2295-2301.

Tada, R and Siever, R, 1989. Pressure solution during diagenesis. *Ann. Rev. Earth Planet. Sci.* **17**, 89-118.

Tuttle, O.F., 1969. Two pressure vessels for silicate-water studies. *Geol. Soc. Am. Bull.* **60**, 1727-1729.

Vidal, O., 1997. Experimental study of the stability of pyrophyllite, paragonite and sodic clays in a thermal gradient. *Eur.J. Mineral.* **9**, 123-140.

Vidal, O. and Durin, L., 1999. Aluminium mass transfer and diffusion in water at 400-550°C, 2 kbar in the $K_2O-Al_2O_3-SiO_2-H_2O$ system driven by a thermal gradient or by a variation of temperature with time. *Minerl. Mag.* **63**(5), 633-647.

Wang,A., Jolliff, B.L. and Haskin L.A., 1999. Raman spectroscopic characterization of a highly weathered basalt: igneous mineralogy, alteration products, and a microorganism, *J. Geophys. Res.* **104**, 27067 -27077.

Weyl, P.K., 1959. Pressure solution and the force of crystallization. A phenomenological theory. *J. Geophys. Res*. **69**, 2001-2025.

Xiao, X., Wirth, R. and Dresen, G., 2002. Diffusion creep in anorthite-quartz aggregates. *J. Geophys. Res.* **107**(B11), 2279-2293.

Yasuhara, H., Elsworth, D. and Polak, A., 2003. A mechanistic model for compaction of granular aggregates moderated by pressure solution. *J. Geophys. Res.* **108**, 2530, doi:10.1029/2003JB002536.

Zulauf, G., Dörr, W., Fiala, J., Koktová, J., Maluski, H. and Valverde-Vaquero, P., 2002. Evidence for high-temperature diffusional creep preserved by rapid cooling of lower crust (North Bohemian shear zone, Czech Republic). *Terra Nova* **14**, 343-354.

Zulauf, J. and Zulauf, G., 2004. Rheology of plasticine used as rock analogue: the impact of temperature, composition and strain.. *J. Struct. Geol.* **26**, 725-737.

**Figures**

Figure 1: Comparison of the viscosity of several kinds of glass versus temperature. Some characteristic temperatures are also reported:
- melting temperature ($T_m$, $\eta = 10$ Pa.s) below which glass is a liquid.
- working range ($\eta = 10^3$-$10^4$ Pa.s) for which glass is easily deformed.
- softening temperature ($\eta = 4 \times 10^6$ Pa.s) which is the maximum temperature for which a glass sample does not flow under its own load.
- glass transition temperature ($T_g$, $\eta = 2 \times 10^{11}$ Pa.s) corresponds to the transition from supercooled liquid to solid glass.
- annealing temperature ($\eta = 10^{12}$ Pa.s) at which glass releases its internal stress (by solid diffusion).
- strain temperature ($\eta = 3 \times 10^{13}$ Pa.s) below which brittle deformation occurs before plastic deformation.

Under the experimental conditions of the present study (T < 450°C) the viscosity of soda-lime glass used (bold curve) is higher than $10^{15}$ Pa.s, thus indicating that elastic brittle deformation occurred in addition to viscous flow (modified from a document from a lecture by L.V. Zhigilei at University of Virginia: http://www.people.virginia.edu/~lz2n/mse209/Chapter14.pdf).

Figure 2: Experimental set-up: A) Sample of glass spheres and fluid with controlled composition encapsulated within a 2mm internal diameter and 20 mm long gold capsule (thickness of the walls: 0.2 mm). The confining pressure medium was water. B) The cold pressure vessel is located within a horizontal heating system. The confining pressure is controlled with water.

Figure 3: SEM-pictures showing the contact morphology in a glass sphere aggregate in experiments performed at 200 MPa and 350°C. A. Flat contacts in dry experiments (Gm2). In experiments with 1 vol.% fluid, three types of contacts are observed: B. Contacts with bonding at the core and a flat rim (GmW3), C. Concentric rims of outgrowths and depressions (GmW12), and D. Stylolite-like contacts (GmW12). E. Filled porosity in 5 vol.% fluid experiments; a similar structure is also observed at 450°C with 1 vol.% fluid (GmW4). F. Filled porosity in 5 vol.% fluid experiments. Microporous recrystallized material, interpreted as Ca-rich zeolites, completely fill the porosity (GmW9). G. Filled porosity in 14 vol.% fluid experiments (ENS2). Increasing the amount of water increases

the degree of reactivity and recrystallization of the spheres. With 15 vol.% fluid (G), the spheres are entirely recrystallized, whereas with 5 vol.% fluid (E, F) recrystallization only occurs at the rims and in 1 vol.% fluid experiments (B, C, D, H) no recrystallization is evidenced. H. Initiation of fractures (arrows) that radiate from one contact to the others (GmW16).

Figure 4: Raman spectra within the glass spheres of experiments performed at 200 MPa and 350°C. a) Spectra of the water content within the material. b) Spectra of the glass spheres and recrystallized material. A. Overview of the difference between experiments: in black an initial dry glass sphere (starting material), in grey a glass sphere from 1 vol.% fluid experiment (GmW1), and in light grey recrystallized material from the porosity of 5 vol.% fluid experiment (GmW4). B. Profiles over a distance of 32 μm from within the glass spheres (bottom) through the recrystallized material (top) in the porosity of an experiment with 5 vol.% fluid (GmW4). The average spacing between successive measurement points is about 4 μm. On all the graphs, the different spectra have been shifted vertically for more clarity.

Figure 5: Raman spectrum of a Ca-bearing Zeolite (thomsonite, from Wang *et al.* 1999).

Figure 6: A) Sketch to illustrate the method of calculating deformation at the grain contact using the initial sphere diameter ($d_s$) and the diameter of the truncated contact ($d_c$). B. Time-dependent change in the average 1-D strain for dry (empty symbols) and 1vol.% fluid (full symbols) experiments performed at 150°C and 350°C (P = 200 MPa, circles and squares respectively). The curves represent mean-square regression fits of the experimental data following a power law. R is defined as the correlation coefficient using: $R = \dfrac{\text{cov}(X,Y)}{\sigma_X \sigma_Y}$, and $\text{cov}(X,Y) = \dfrac{1}{n}\sum_{j=1}^{n}(X_j - \overline{X})(Y_j - \overline{Y})$ where $\overline{X}$ and $\overline{Y}$ are the average values of X and Y and $\sigma_X$ and $\sigma_Y$ are the standard deviations.

Figure 7: Change in the average contact deformation of dry (grey diamonds) and 1 vol.% fluid (black diamonds) experiments versus temperature (P = 200 MPa). The full lines represent mean-square regression fits of the experimental data using an exponential law. Error bars represent the standard-deviation normalized to the average strain. R is defined as in figure 6.

Figure 8: Change in the average contact deformation of dry (grey diamonds) and 1 vol.% fluid (black diamonds) experiments versus confining pressure (T = 350°C). The full lines represent fits mean-square regression of the experimental data using a linear law. Error bars represent the standard-deviation normalized to the average strain. R is defined as in figure 6.

Figure 9: Evidence of dissolution and indentation at the contact between glass spheres and mica from SEM-pictures. A) Mica mechanical indentation in a dry experiment (Gm2). B) Kinked mica and broken spheres in a 1 vol.% experiment (GmW2). C) Dissolution of a glass sphere at the contact with a mica flake in experiment with 1 vol.% water (black arrow, GmW10). D) Dissolution of a glass sphere at the contact with a mica flake in an experiment with 5 vol.% water (white arrow, GmW11).

Figure 10: Dissolution patterns on quartz grains (QmW10). A) Morphology of a contact between two quartz grains showing evidence of intergranular pressure solution (micro-stylolite). The small debris are produced by grain crushing under brittle deformation. B) View of the quartz aggregate and the previous contact.

Figure 11: Strain rate ($s^{-1}$) versus contact (local) stress (GPa) for both dry (empty symbols) and 1 vol.% experiments (full symbols) performed at 150°C and 350°C (circles and squares respectively) using an exponential deformation law ($ln\,\dot{\varepsilon}$ versus $\sigma$, from equation 5). The black lines represent exponential fits of all experiments assuming a constant slope for $B = 56$ Å$^3$ (see text for more explanations). The full lines represent fits of 1 vol.% fluid experiments at 350°C, the dashed line the fit of dry experiments at 350, and the dotted line the fit of both dry and 1vol.% fluid experiments at 150. The grey line is the best exponential fit of the experiments performed with 1vol.% fluid ($B = 69$ Å$^3$). The fits and regression equations are given in SI units ($\sigma$ in Pa). R is defined as in figure 6.

Figure 12: Strain rate ($s^{-1}$) versus stress (Pa) for both dry and 1 vol.% experiments performed at 150°C and 350°C following a power law ($ln\,\dot{\varepsilon}$ versus $ln\,\sigma$, from equation 6). The full and dashed lines represent the best exponential fits of 1 vol.% fluid experiments and dry experiments, respectively. These two lines have the same slope (n = 10.5). Experiments performed at 150°C are fitted using n = 10.5 as for the 350°C experiments. Dry and 1vol.%

856   experiments are fitted with the same power law. The fits and regression equations are given
857   in SI units ($\sigma$ in Pa). R is defined as in figure 6.
858
859   **Tables**
860   Table 1: Data of experiments performed with glass spheres and quartz grains.
861   Table 2: Deformation law parameters.

**Figure 1**
**Click here to download high resolution image**

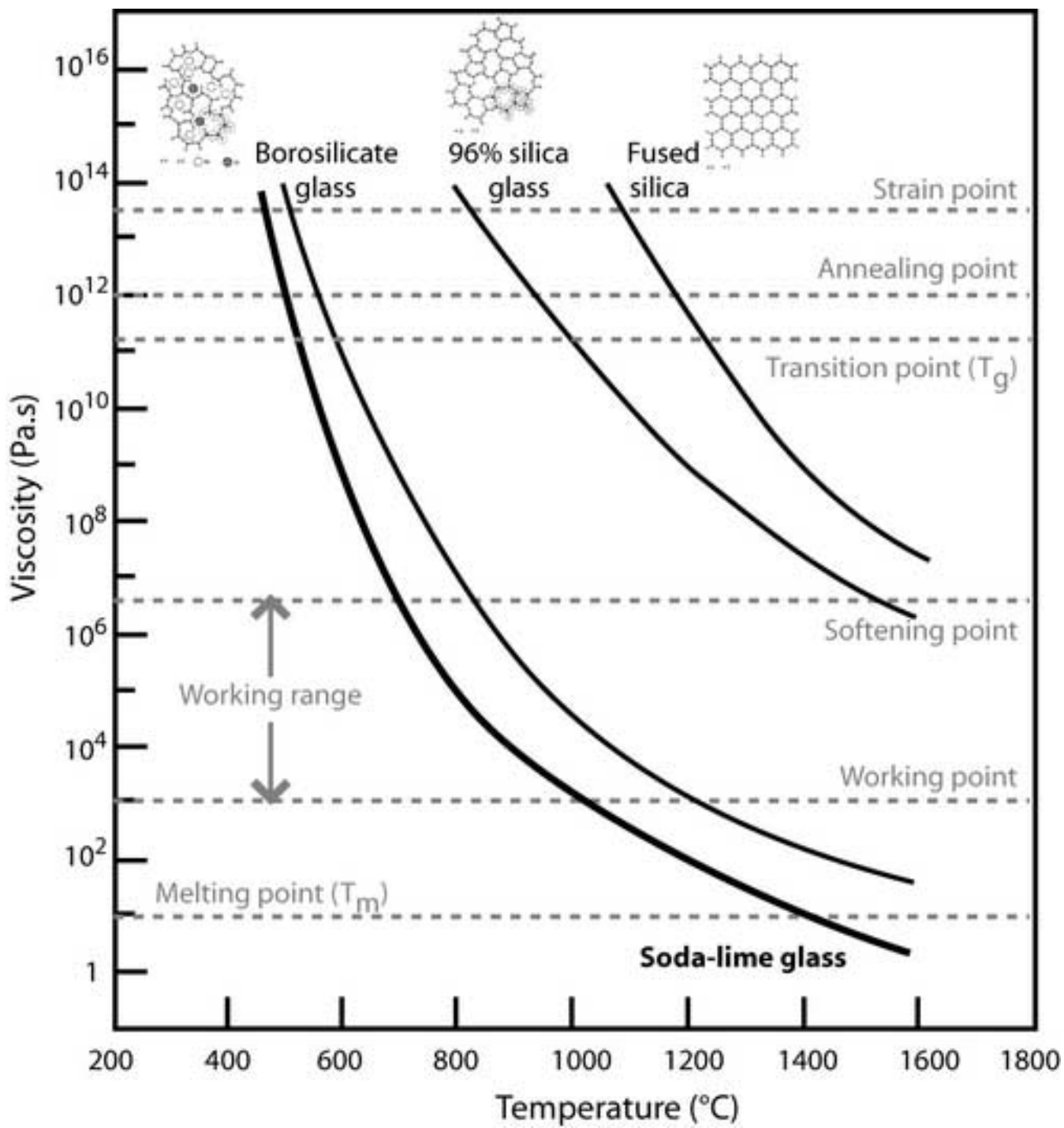

**Figure 2**
**Click here to download high resolution image**

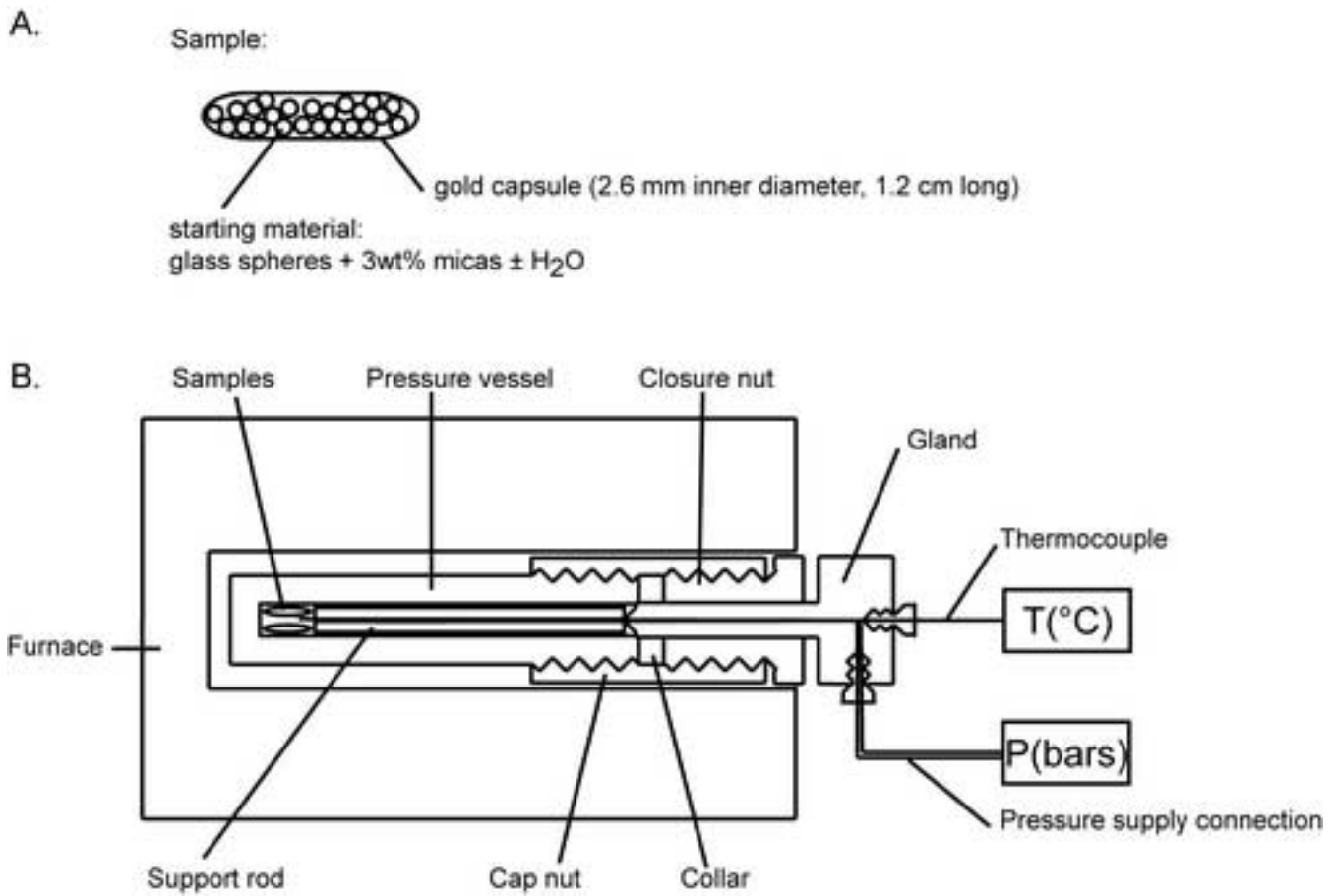

**Figure 3**
Click here to download high resolution image

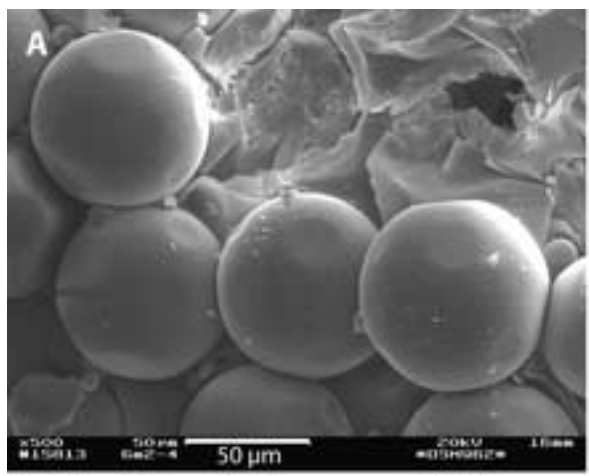
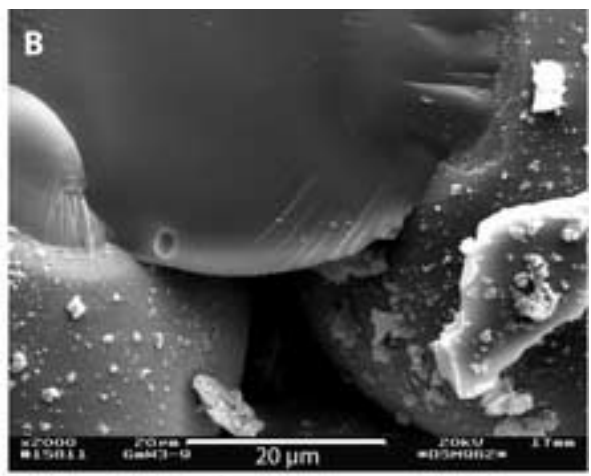
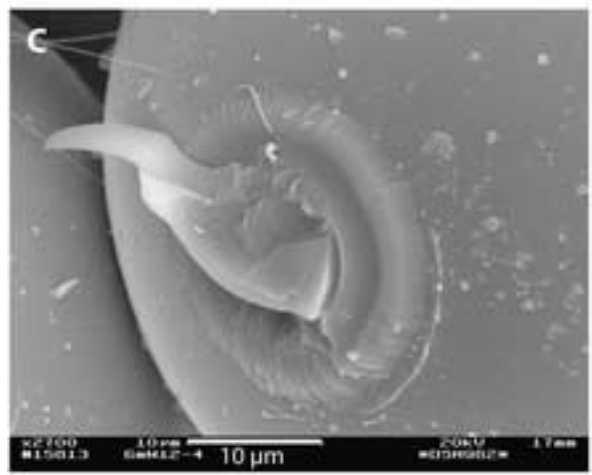
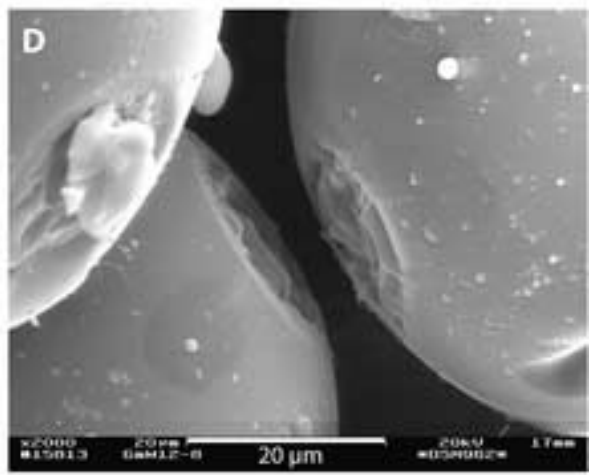
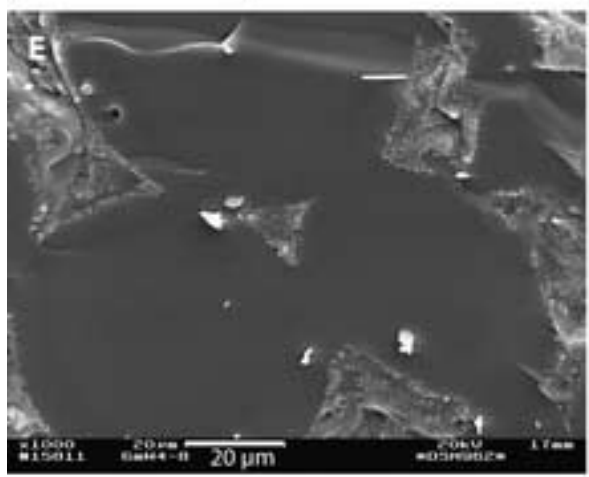
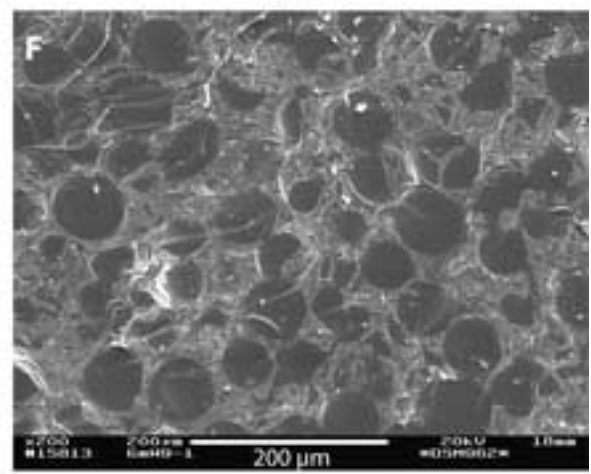
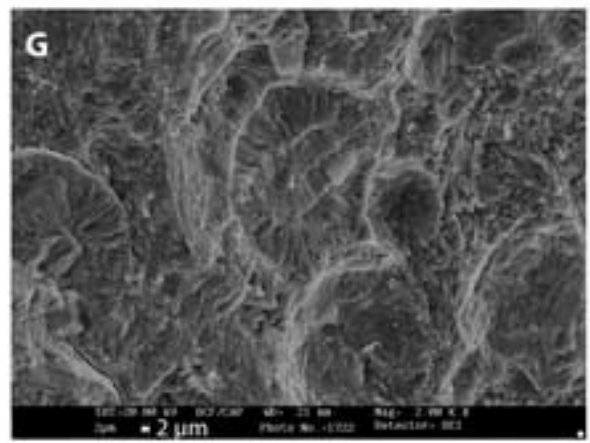
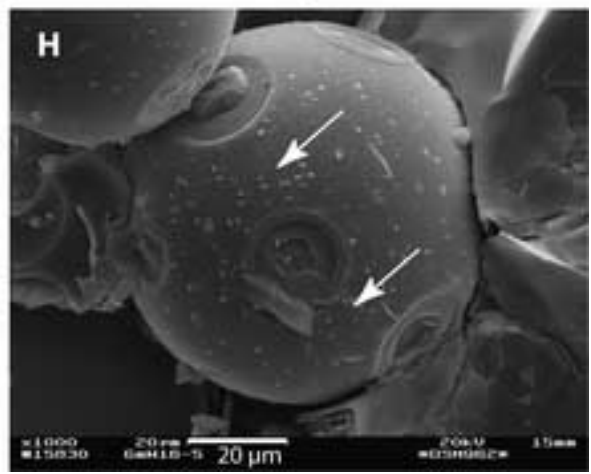

**Figure 4**
**Click here to download high resolution image**

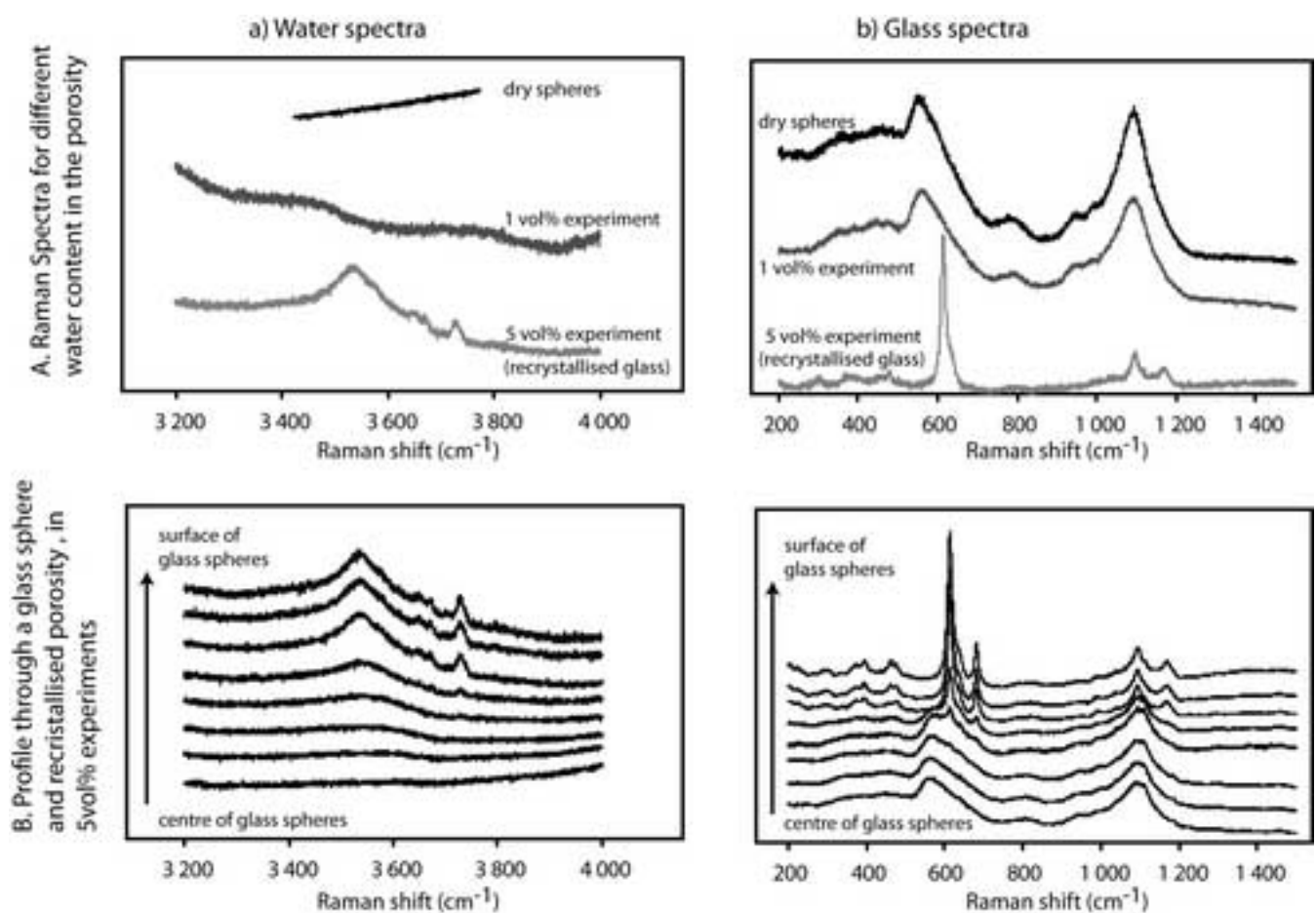

**Figure 5**
**Click here to download high resolution image**

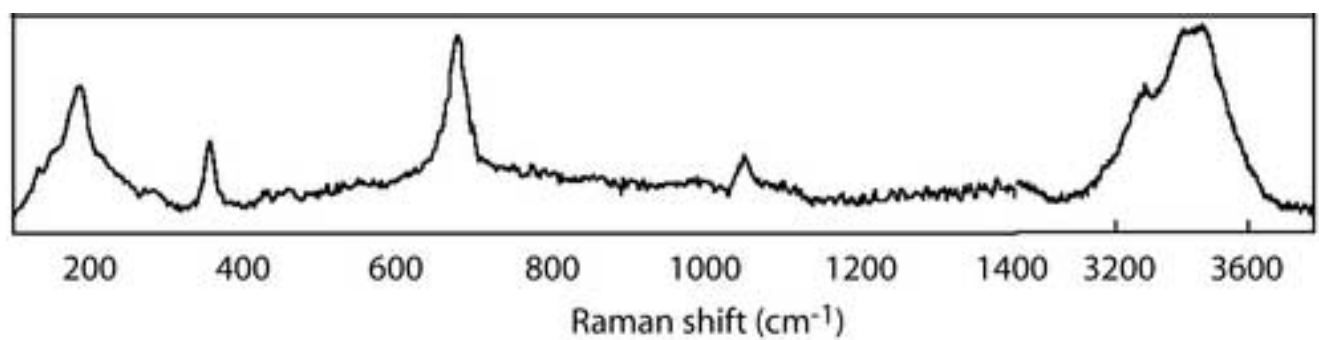

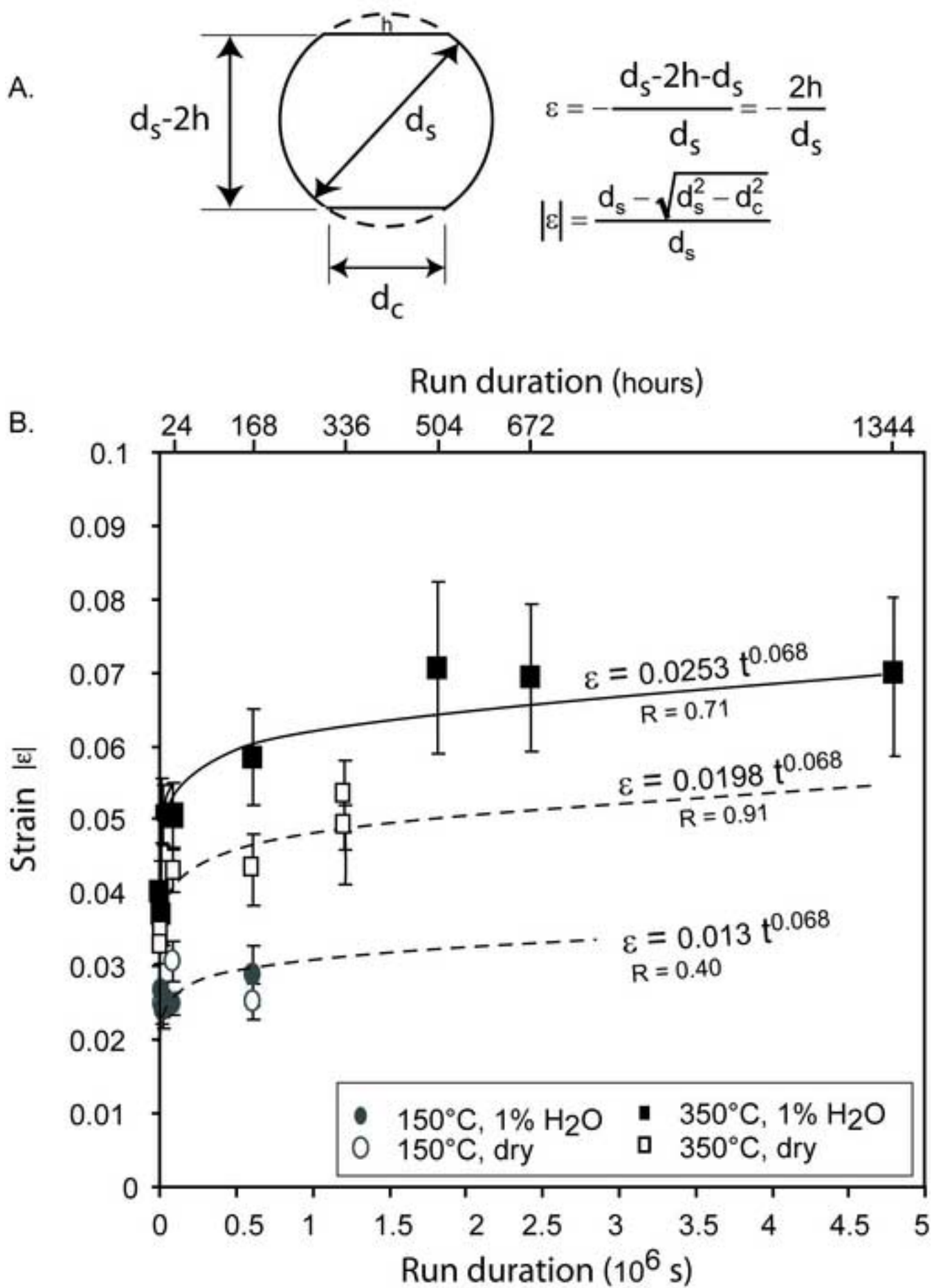

**Figure 7**
Click here to download high resolution image

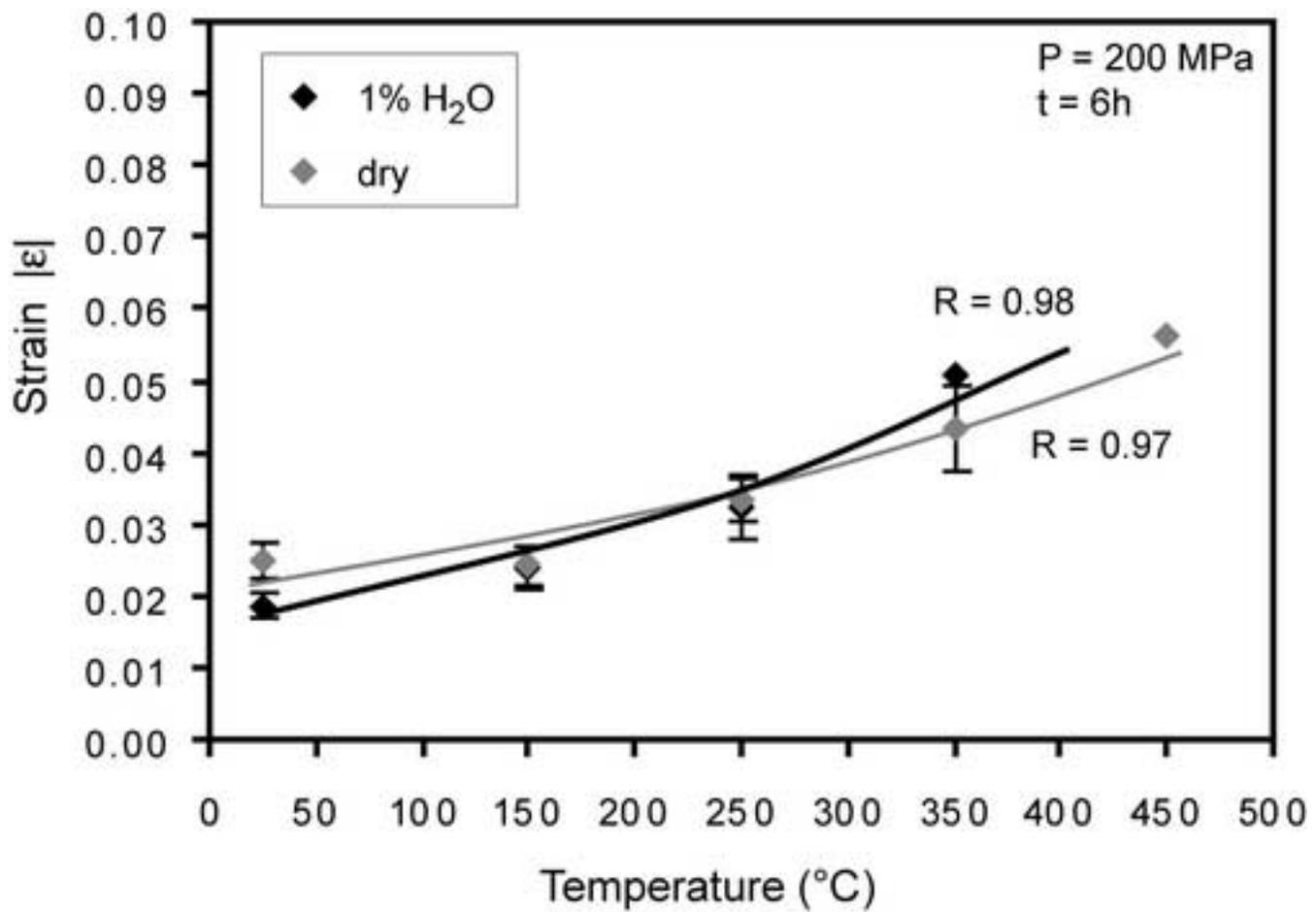

**Figure 8**
**Click here to download high resolution image**

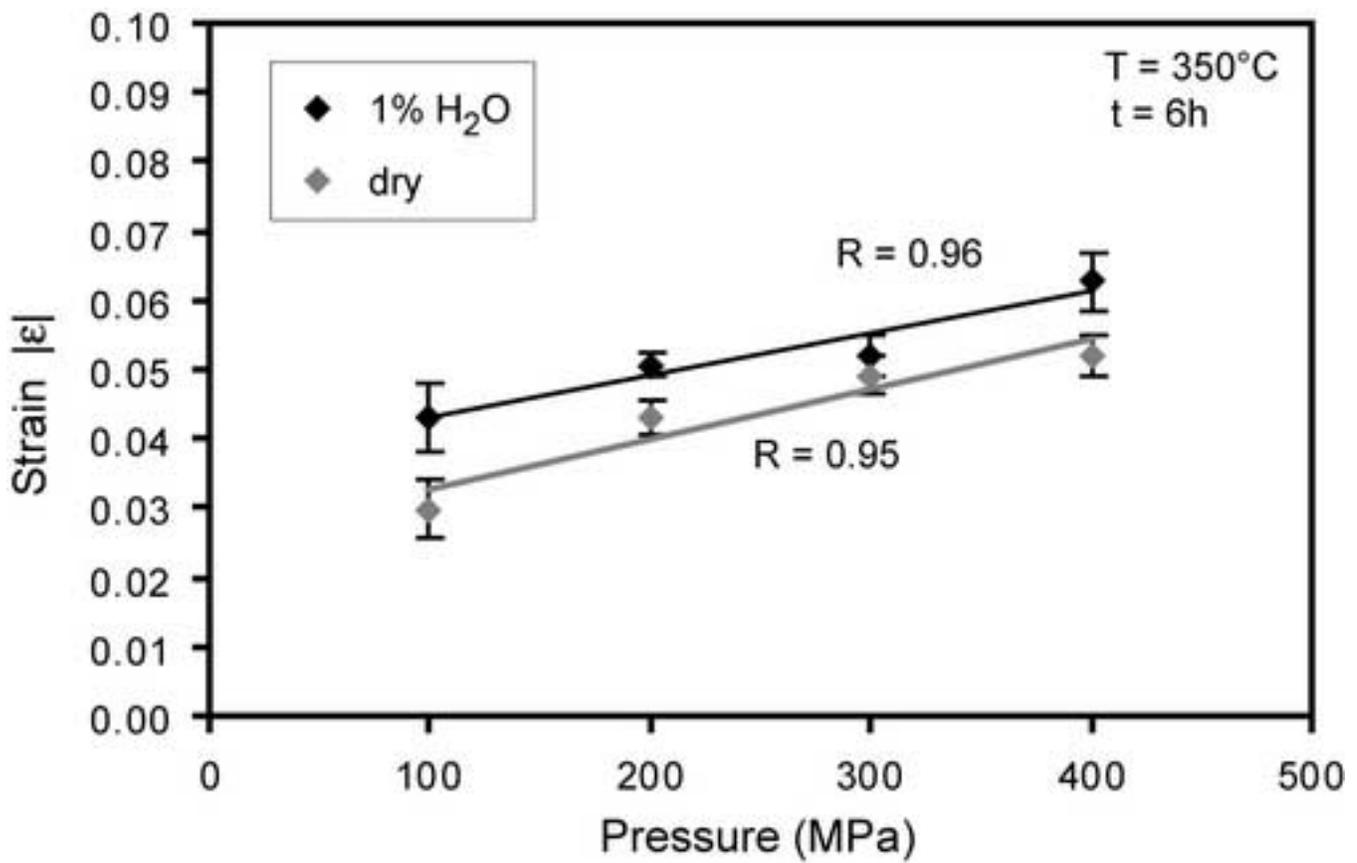

**Figure 9**
**Click here to download high resolution image**

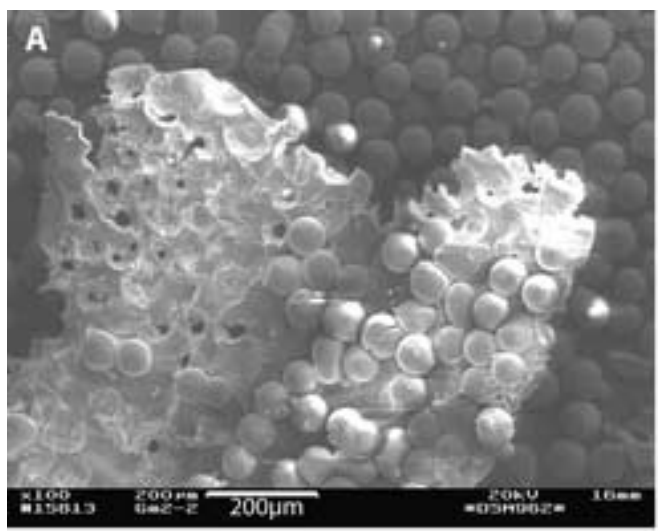
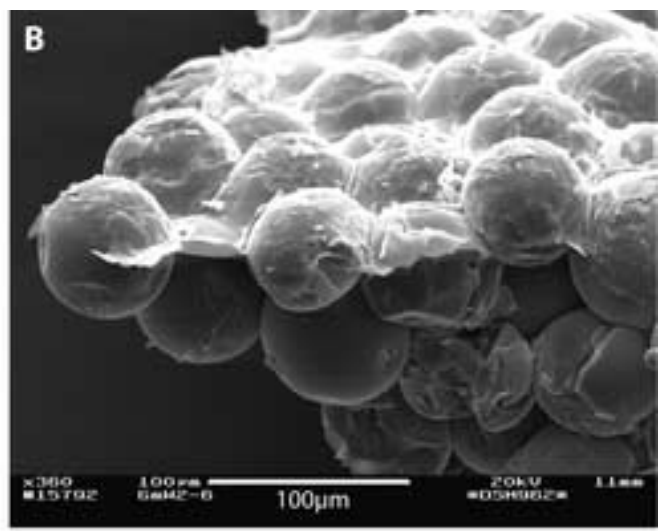
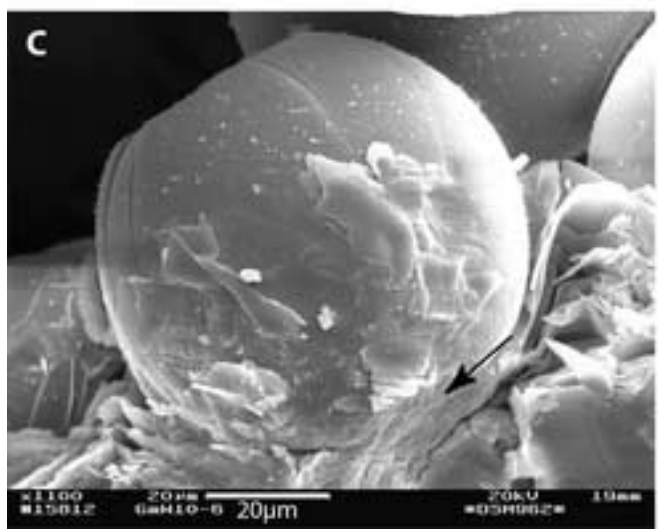
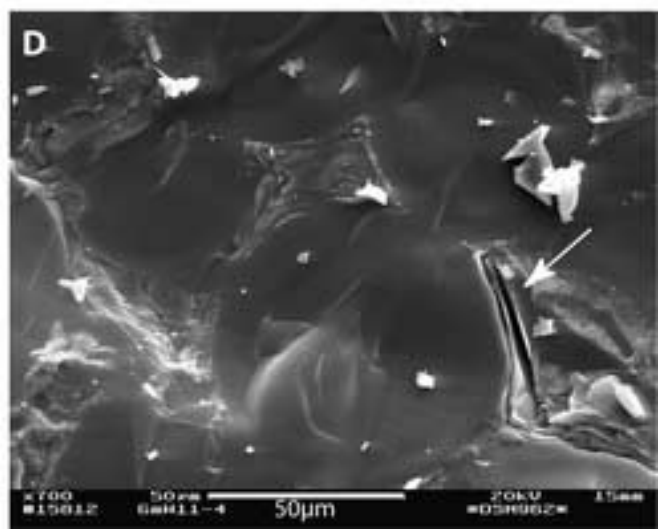

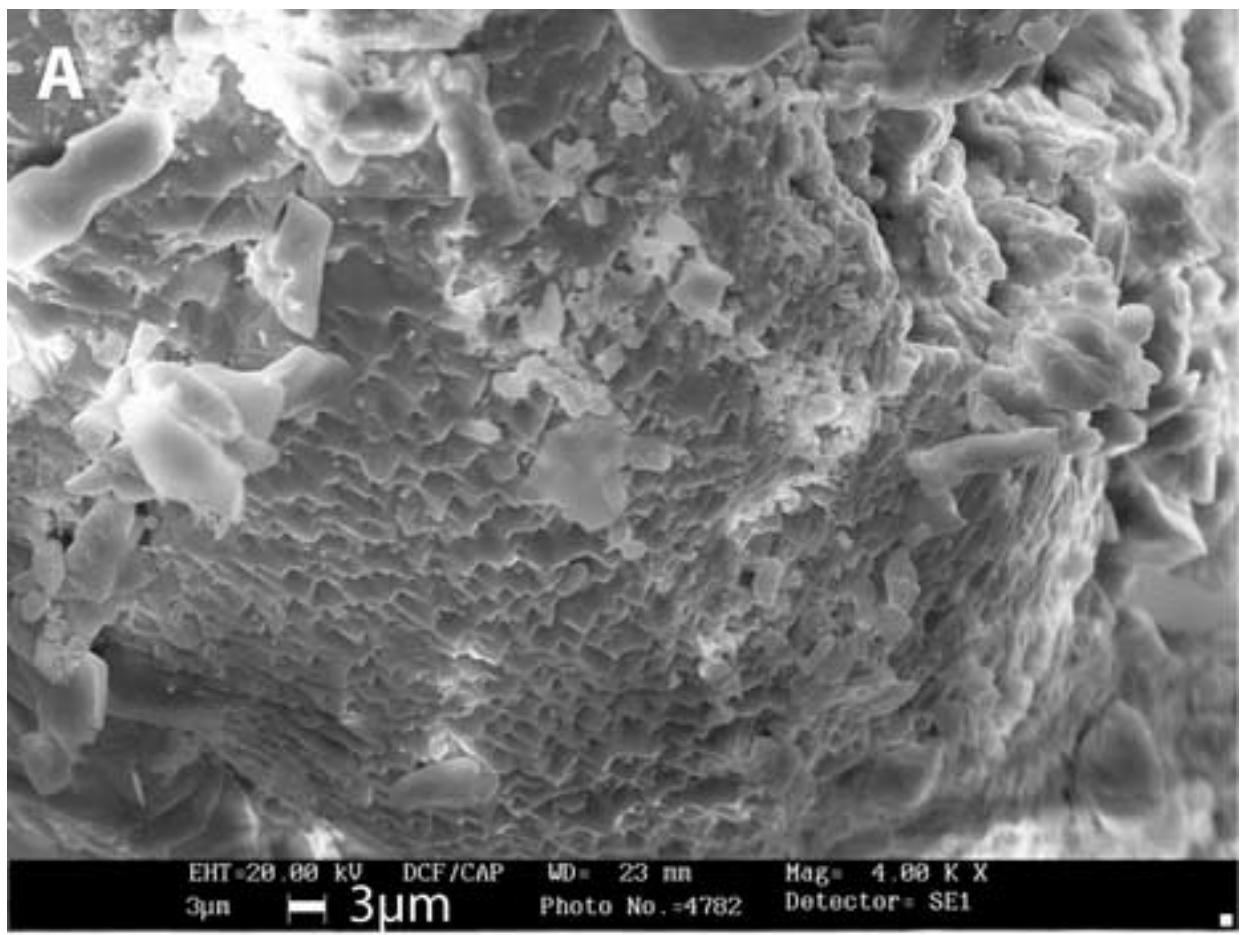
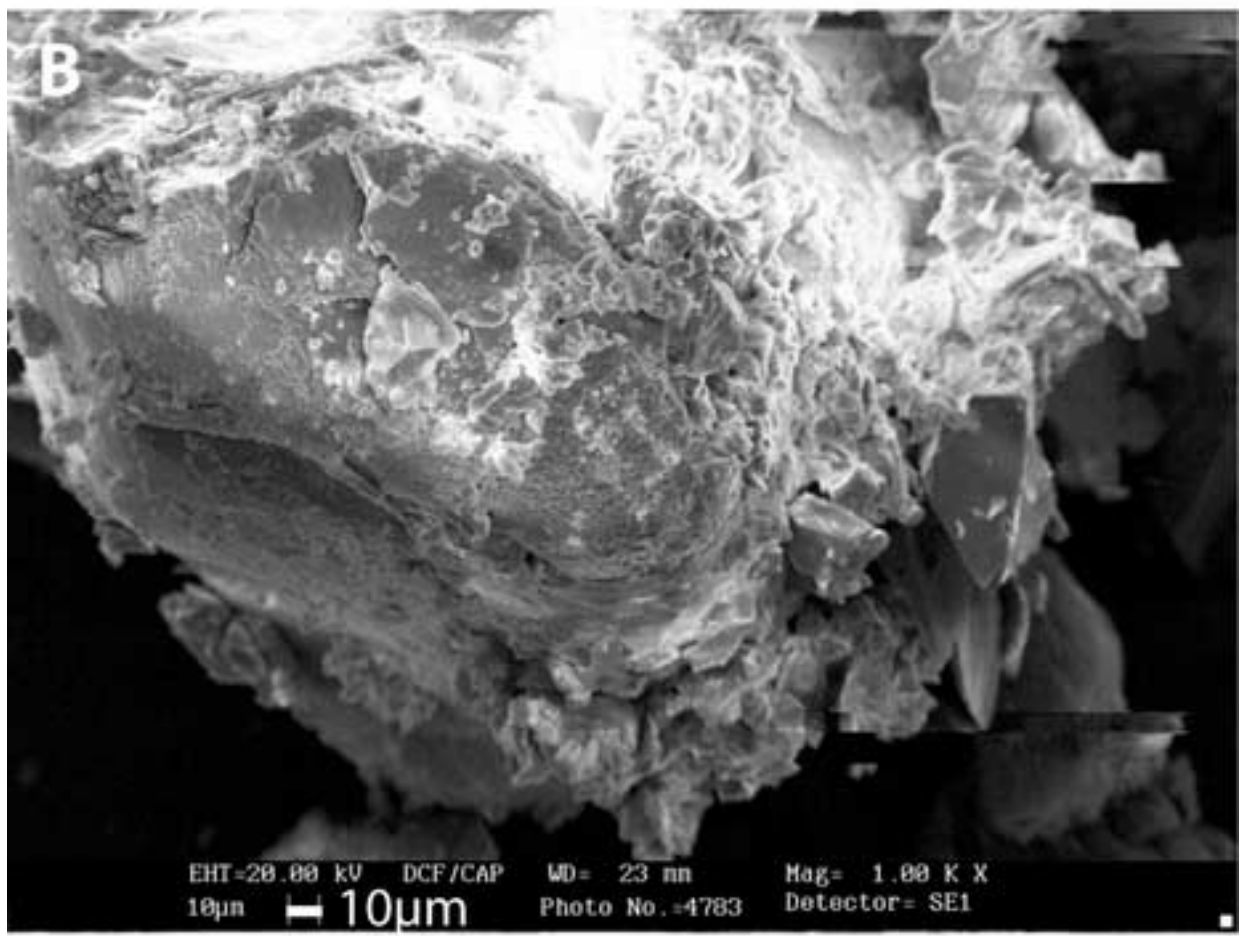

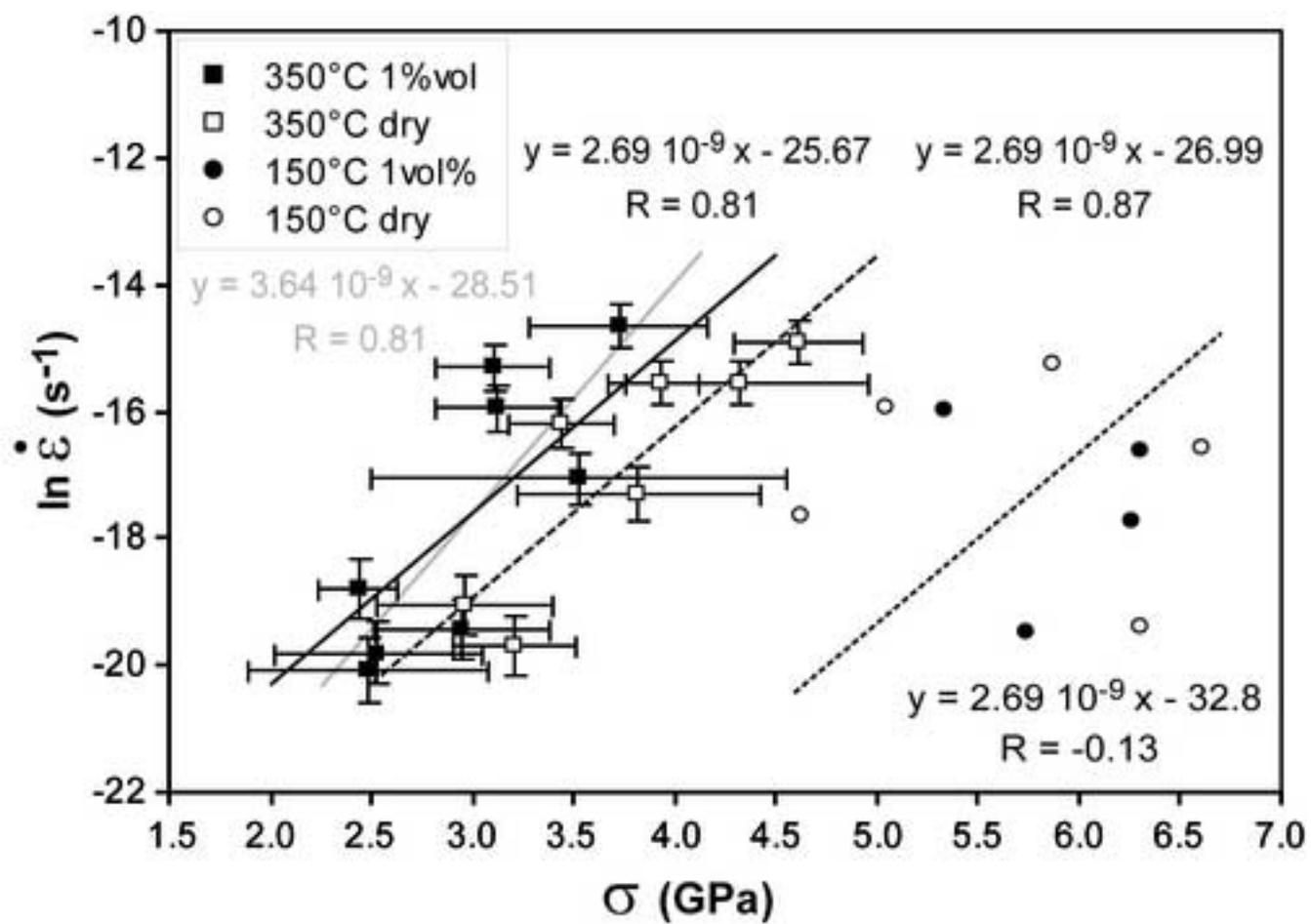

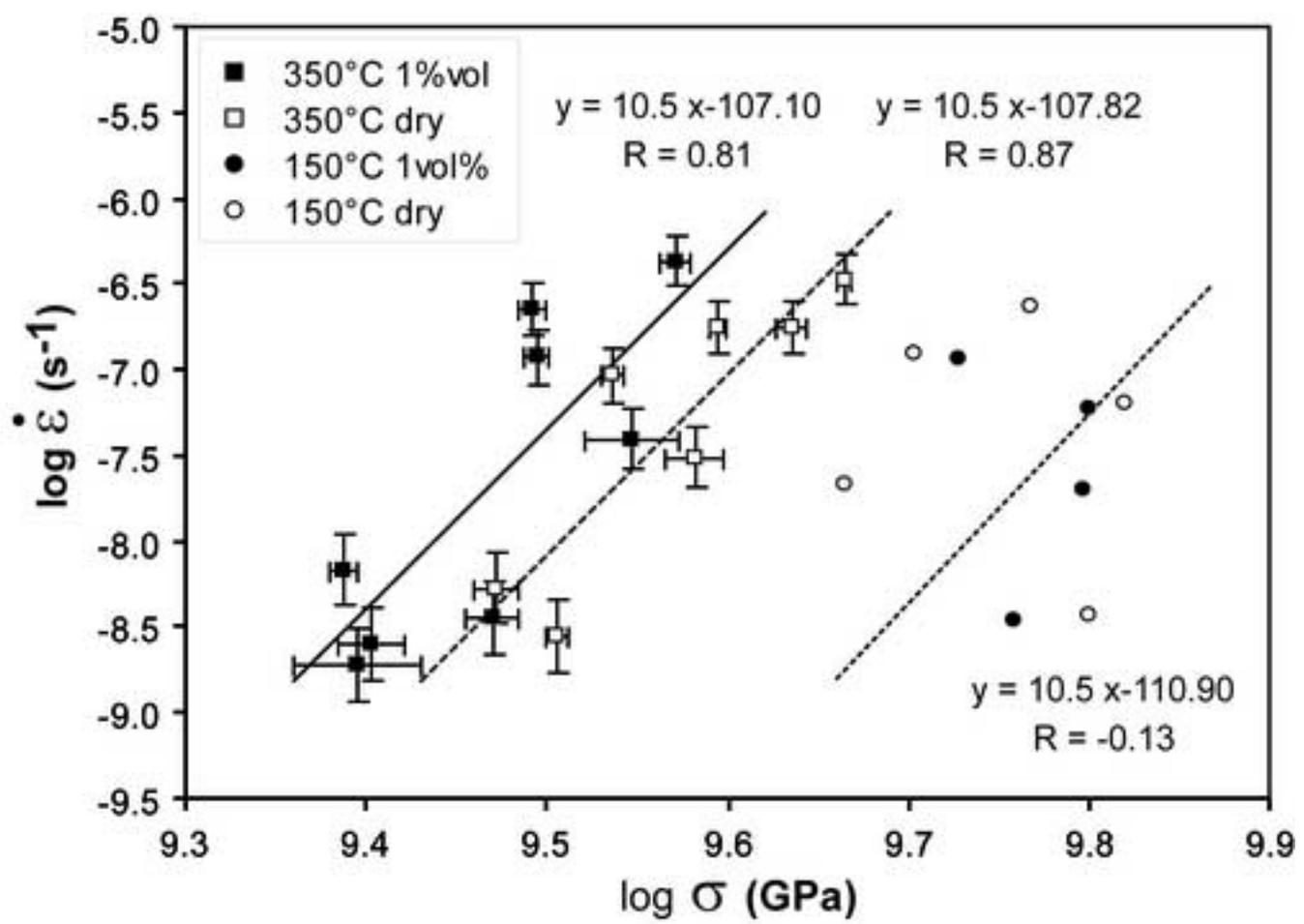